\documentclass[aps,showpacs,superscriptaddress]{revtex4}
\usepackage{amsmath,amssymb,graphicx,afterpage,here}
\usepackage{afterpage,subfigure,epsfig,psfig,rotating,minitoc}
\usepackage{ae}
\usepackage{dcolumn}
\newcommand{\be}{\begin{equation}}
\newcommand{\ee}{\end{equation}}
\newcommand{\bea}{\begin{eqnarray}}
\newcommand{\eea}{\end{eqnarray}}
\newcommand{\nab}{{\mbox{\boldmath $\nabla$}}}
\newcommand{\parti}[2]{\frac{\partial #1}{\partial #2}}

\newcommand{\ub}{{\bf u}}

\newcommand{\noi}{\noindent}

\newcommand{\bb}[1]{{\mbox{\boldmath $#1$}}}

\begin{document}

\title{Convective  instabilities in two superposed horizontal liquid layers heated laterally\\}
\author{S. Madruga }
 \email{smadruga@fisica.unav.es}
\affiliation{Departement AGO, Li\`ege University, B5, Sart-Tilman, B-4000 Li\`ege, Belgium}
 \affiliation{Instituto de F\'{\i}sica,  Universidad de Navarra, E-31080 Pamplona, Navarra, Spain}
\author{C. P{\'e}rez--Garc\'{\i}a}
\affiliation{Instituto de F\'{\i}sica,  Universidad de Navarra,
E-31080 Pamplona, Navarra, Spain}
 \affiliation{ Depto.
Ingenier\'{\i}a Mec\'anica,  E. T. S.  Ingenieros
(TECNUN),Universidad de Navarra,  E-20018 San Sebasti\'an, Spain}
\author{G. Lebon}
\affiliation{Departement AGO, Li\`ege University, B5, Sart-Tilman, B-4000 Li\`ege, Belgium}

\date{\today}
\begin{abstract}

This work is devoted to the theoretical study of the stability of two superposed horizontal
liquid layers bounded by two solid planes and subjected to a horizontal temperature gradient.
 The liquids are supposed to be immiscible with a nondeformable  interface.
 The forces acting on the system are buoyancy and interfacial tension.
Four different flow patterns and temperature profiles
are found for the basic state.  A linear perturbative analysis with respect to two and three
dimensional perturbations reveals the existence of three kind of patterns.
Depending on the relative height of both liquids several situations are predicted:
either wave propagation from cold to the hot regions,
or waves propagating in the opposite direction or still stationary longitudinal rolls.
The behavior of three different pairs of liquids which have been used in
experiments on bilayers under vertical gradient by other authors have been examined. The
instability mechanisms are discussed and a  qualitative interpretation of the different behaviors exhibited by
the system is provided.  In some configurations it is
possible to find a codimension-two  point created by the interaction of two Hopf modes with different frequencies
 and wavenumbers. These results suggest to consider two liquid layers as an
interesting prototype for the study of propagation
and interaction of waves in the context of the B\'enard-Marangoni problem.
\end{abstract}


\maketitle

\section{Introduction}
There has been recently an increasing interest in the study of buoyant-thermocapillary flows induced by horizontal
thermal gradients \cite{Schatz}. Such situations occur in oceanographic currents due to non homogeneous
heating,
and their interplay with the atmosphere gives rise to complex meteorological phenomena. They are also important in crystal growth. The floating zone process generates thermocapillary
convective cells in the region of crystallization which require better understanding and control.

  When a horizontal temperature gradient is acting on a horizontal liquid layer
enclosed between two horizontal walls the problem is much more complicated than in the
the more classical case of vertical heating.
First, the reference state is no longer at rest,  giving rise to a general flow and a non-linear vertical temperature profile. A non trivial problem
will therefore consist in determining the basic reference temperature and velocity fields.
Another peculiarity is that the
threshold of instability depends on the Prandtl number $Pr$.
  The problem of thermocapillary instabilities induced by a lateral heating was first studied
 by Smith and Davis \cite{SmithDavis}. They predicted the presence of hydrothermal waves and
stationary rolls. More details are provided in a subsequent work by
Smith \cite{Smith1}, who showed the existence of two different mechanisms of instability depending on
$Pr$. At low $Pr$, the energy necessary to sustain the disturbances comes from the
horizontal applied temperature field and the hydrothermal waves propagate in a direction
perpendicular to the horizontal temperature gradient. At high $Pr$, the energy is extracted from the
vertical temperature field by vertical convection and it gives rise to hydrothermal waves propagating parallel 
to the
temperature gradient. At intermediate $Pr$, the mechanism is a
combination of the previous effects and leads to waves forming an angle with the streamwise direction.
Parmentier et al. \cite{Parmentier} studied the coupling of buoyant and thermocapillary driven
instabilities, in the case of systems with
lower and upper insulating bounding surfaces within a 3D
linear formulation. Their numerical results
 display the three kinds of behavior described above.
For thermally conducting surfaces, Gershuni et al. \cite{Gershuni} find stationary rolls for $Pr>1$. Other theoretical works have  mainly focused on comparison with experiments. Mercier and Normand \cite{Mercier} 
performed a linear analysis to explain the experimental results by Daviaud and Vince \cite{Daviaud} for a
silicon oil with $Pr =10$. They notice that the transition between traveling waves and stationary rolls 
observed
when the liquid depth is increased depend on the heat transfer between the liquid and its
environment. In their discussion about absolute, global and
convective instabilities Priede and Gerbeth \cite{Priede} show that the threshold of absolute instability fits
better the experimental data of $1\; cSt$ silicon oil ($Pr =13.9$) \cite{Riley} than the
value predicted from
convective instability.
 A non-linear analysis of stability for purely thermocapillary convection has been carried out by Smith \cite{Smith2},
who determines the range of existence of the two oblique traveling waves predicted by the linear theory.
  More recently, new results have been reported \cite{Pela-Burg}  on
hydrothermal waves in different geometries, measuring their frequency, wavenumber and angle of propagation.

In the previous theoretical works it is assumed that the  liquid is surmounted by a gas phase above that 
remains passive (one-layer approximation).
In some cases \cite{Mercier,Priede}
the effect of the adjacent gas is modelled by a Biot number, a phenomenological parameter that characterizes the
heat transfer between liquid and gas.  The Biot number was originally defined for steady conditions; however
when convection takes place, the dynamics of thermal and mechanical perturbations in the gas may
have a strong
influence on the convection inside the liquid. In such a case, the introduction of a Biot number becomes questionable
 and the one-layer approximation should be revised. In the general case, a full two-layer hydrodynamical
description is required. Furthermore well controlled experiments involve
always two fluid layers \cite{Tok}.
 Such systems are typical of many practical situations as Earth's mantle convection
  or encapsulated crystal growth.

 In the past, most theoretical \cite{Engel} and experimental studies
 \cite{And}-\cite{Car} about Rayleigh-B\'enard convection in two immiscible liquids have been
devoted to  vertical heating.
  Only few contributions are found dealing with a horizontal temperature gradient.
To our knowledge the first work on the subject is by Villers and Platten \cite{Vill-Plat}; they measured
the velocity profiles in each layer as a function of the height in a system formed by water and heptanol. They
also developed  a simple theoretical model to calculate the horizontal velocity profile
as a function of the ratio of viscosities, expansion coefficients and
thicknesses of both layers. Later on, Doi and Koster \cite{DoiKoster} studied
theoretically the thermocapillary convection under microgravity conditions in two
immiscible liquid layers with a free upper surface, in 2-D geometries. They determine under which conditions the
lower layer remains at rest. Moreover they carry out numerical simulations in a box of aspect ratio $4$ in 
order to analyze the effect of the vertical walls.
Numerical simulations in cavities of different aspect ratios for coupled thermocapillary and
buoyancy-driven convection
were performed by Liu {\em et al.}\cite{Liu-Rou}, and an asymptotic solution for the velocity
 in the limit of infinite aspect ratio has also been derived \cite{Liu-Rou2}.

The objective of the present work is to study
coupled thermocapillary and buoyancy convection in a two-layer
system of infinite horizontal extension subject to lateral heating
and to find out the regimes occurring in this configuration. A linear 
approach to the problem is carried out, showing
the kind of oscillatory and stationary behaviors that occur
 in this system.

The paper is organized as follows:  In section
\ref{sec:problemdefinition}, the general equations governing the
problem  are established. Section \ref{sec.basicstate} is devoted
to the derivation of the temperature and velocity profiles of the
basic flow, the  explicit expressions are reported in
the Appendix. In section \ref{sec.evolequation} is carried out a
linear stability analysis of the basic state towards bidimensional
and three dimensional perturbations. Conclusions are drawn in
Section \ref{sec.conclusions}.

\section{Problem formulation
\label{sec:problemdefinition}}

\begin{center}
\begin{figure}[t]
\centering
\includegraphics[width=\textwidth]{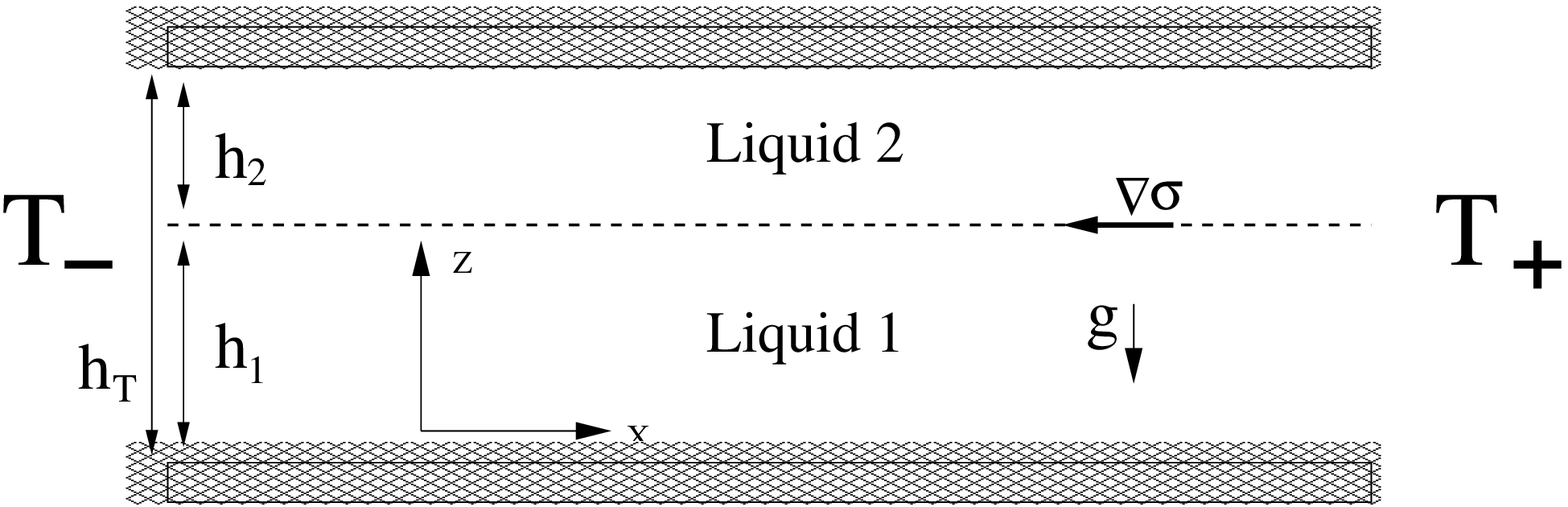}
\caption{
Diagram of the two-layer system under study. Two superposed immiscible liquids
 bounded horizontally by conductive rigid walls are subject to a
 horizontal gradient of temperature. The interface is supposed to be non
 deformable. The gravity and interfacial tension are the forces acting on
the system.
\label{f:esquema}}
\end{figure}
\end{center}

 We consider a system of two horizontal superimposed immiscible liquid layers  of thickness $h^{(i)}$,
densities $\rho^{(i)}$, kinematic viscosities $\nu^{(i)}$, coefficients of volume expansion $\alpha^{(i)}$,
thermal conductivities $\lambda^{(i)}$ and thermal diffusivities $\kappa^{(i)}$, superscript $i=1,2$ refers to the
lower and upper fluid respectively. The system is infinite in the horizontal extension and is limited in the
vertical direction by two horizontal rigid and thermally conductive walls (see
Fig. \ref{f:esquema}). The interface between both liquids is
assumed to be horizontal and nondeformable. The origin of the Cartesian reference system is fixed at the 
interface.

 The system is heated, with a constant temperature
gradient $\beta$ imposed along the horizontal direction,  which produces a conducting temperature profile 
 given by
$T=T_-+\beta x$,
where $T_-$ is the temperature of the cold side. The Boussinesq approximation is taken for granted in both layers. The system is subject
to the gravity field and its density is given by the state equation $\rho^{(i)}=\rho_0^{(i)}[1-\alpha^{(i)}(T-T_-
)]$. The interfacial tension which acts at the interface admits a temperature variation approximated
by the linear equation of state $\sigma=\sigma_0-\gamma (T-T_-)$ where
$\gamma=-\frac{\partial\sigma}{\partial
T}$ is a positive constant (at least for the pairs of liquids considered in this work).

  The governing equations expressing the balance of mass, momentum and energy in the Boussinesq approximation
are given by

\begin{subequations}\label{eq:nd1}
\begin{eqnarray}
\!\!\!\!\!\!\!\!\!\!\nab\cdot\ub^{(i)}&=&0, \\
\!\!\!\!\!\!\!\!\!\!\parti{\ub^{(i)}}{t}+(\ub^{(i)}\cdot\nab)\ub^{(i)}&=&-\frac{\nab p^{(i)}}{\rho_{0}^{(i)}}-g[1-\alpha(T^{(i)}-T_-)]\hat z+  \nu^{(i)}\nabla^{2}\ub^{(i)},\\
\!\!\!\!\!\!\!\!\!\!\parti{T^{(i)}}{t}+(\ub^{(i)}\cdot\nab) T^{(i)}&=&\kappa^{(i)}\nabla^{2} T^{(i)},
\end{eqnarray}
\end{subequations}

\noi wherein $\bb{u}^{(i)}(u^{(i)},v^{(i)},w^{(i)})$ is the velocity field, $p^{(i)}$ the
pressure, $g$ the
acceleration of the gravity, and $\hat z$ the unit vector in the vertical direction.
The boundary conditions at the rigid and thermally conductive bottom and upper walls are

\begin{subequations}\label{eq:nd2}
\begin{eqnarray}
z=-h^{(1)} \to \;\;\;\;\;\; \ub^{(1)}=0;\;\;\;\;\; T^{(1)}=T_{w}\;\; (T_{w}=T_-+\beta x),\\
z=h^{(2)} \to \;\;\;\;\;\; \ub^{(2)}=0;\;\;\;\;\; T^{(2)}=T_{w}\;\; (T_{w}=T_-+\beta x).
\end{eqnarray}
\end{subequations}

\noi  Furthermore, one must include the conditions of
continuity of the temperature, heat flow and velocity at the interface (the normal velocity component is zero
because the interface is supposed to be nondeformable)

\begin{eqnarray}\label{eq:ndbc}
z=0\to&& \;\;\;\;\;\; \ub_h^{(1)}=\ub_h^{(2)},\;\;\;\;\;w^{(1)}=w^{(2)}=0,\;\;\;\;T^{(1)}=T^{(2)}, \nonumber\\
&&\lambda^{(1)}\partial_{z}T^{(1)}=\lambda^{(2)}\partial_{z}T^{(2)},\;\;\;\;\;\;\partial_{z}{\ub_{h}^{(2)}}
-\partial_{z}{\ub_{h}^{(1)}}=-\frac{d\sigma}{d T}\nab_{h} T^{(1)},
\end{eqnarray}

\noi the last relation in (\ref{eq:ndbc}) expresses the balance between the
tangential stresses at the interface. To transform the
governing equations and boundary conditions in dimensionless form, the following  scales are selected: for length
$h^{(1)}$, time ${h^{(1)}}^{2}/\kappa^{(1)}$, velocity $\kappa^{(1)}/h^{(1)}$, pressure $\rho^{(1)}\nu^{(1)}
\kappa^{(1)}/{h^{(1)}}^{2}$ and temperature $\beta h^{(1)}$.
In terms of non-dimensional quantities, the balance Eqs. (\ref{eq:nd1}) for the lower liquid
read as

\begin{subequations}\label{eq:ndbi1}
\begin{eqnarray}
\!\!\!\!\!\!\!\!\!\!\nab\cdot\ub^{(1)}&=&0, \\
\!\!\!\!\!\!\!\!\!\!Pr^{-1} [\parti{\ub^{(1)}}{t}+(\ub^{(1)}\cdot\nab)\ub^{(1)}]&=&-\nab p^{(1)}-[\frac{(h^{(1)})^{3} g}{\nu^{(1)}\kappa^{(1)}}-Ra(T^{(1)}-T_-)]\hat z+ \nabla^{2}\ub^{(1)},\\
\!\!\!\!\!\!\!\!\!\!\parti{T^{(1)}}{t}+(\ub^{(1)}\cdot\nab) T^{(1)}&=&\nabla^{2} T^{(1)},
\end{eqnarray}
\end{subequations}

\noindent while for the upper liquid

\begin{subequations}\label{eq:ndbi2}
\begin{eqnarray}
\!\!\!\!\!\!\!\!\!\!\nab\cdot\ub^{(2)}&=&0, \\
\!\!\!\!\!\!\!\!\!\!Pr^{-1} [\parti{\ub^{(2)}}{t}+(\ub^{(2)}\cdot\nab)\ub^{(2)}]&=&-\frac{1}{\rho}\nab p^{(2)}-[\frac{(h^{(1)})^{3} g}{\nu^{(1)}\kappa^{(1)}}-\alpha \,Ra(T^{(2)}-T_-)]\hat z+ \nu\nabla^{2}\ub^{(2)},\\
\!\!\!\!\!\!\!\!\!\!\parti{T^{(2)}}{t}+(\ub^{(2)}\cdot\nab) T^{(2)}&=&\kappa\nabla^{2} T^{(2)}.
\end{eqnarray}
\end{subequations}


\noindent The nondimensional boundary conditions are:

\begin{subequations} \label{eq:bcin}
\begin{eqnarray}
&&\!\!\!\!\!\!\!\!\!\!\!\!z=-1\to\;\;\;\;\, \ub^{(1)}=0,\;\;\;\;\; T^{(1)}=T_{w}\;\; (T_{w}=T_-+x); \hspace{4cm}\label{eq:bcin-a} \\
&&\!\!\!\!\!\!\!\!\!\!\!\!z=a\to \;\;\;\;\;\;\; \ub^{(2)}=0,\;\;\;\;\; T^{(2)}=T_{w}\;\; (T_{w}=T_-+x); \hspace{4cm}\label{eq:bcin-b} \\
&&\!\!\!\!\!\!\!\!\!\!\!\!z=0\to \;\;\;\;\;\;\; \ub^{(1)}=\ub^{(2)},\;\;\;\;\;w^{(1)}=w^{(2)}=0,\;\;\;\;T^{(1)}=T^{(2)}, \\
&&\;\;\;\;\partial_{z}T^{(1)}=\lambda\partial_{z}T^{(2)}, \;\;\;\;\;
\;\mu\partial_{z}{\ub_{h}^{(2)}}-\partial_{z}{\ub_{h}^{(1)}}=Ma\nab_{h}T^{(1)}.\hspace{1cm} \nonumber
\end{eqnarray}
\end{subequations}

\noindent In (\ref{eq:ndbi1}-\ref{eq:bcin}) the following nondimensional parameters  have been introduced
$\alpha=\alpha^{(2)}/\alpha^{(1)}$, $\kappa=\kappa^{(2)}/\kappa^{(1)}$, $\nu=\nu^{(2)}/\nu^{(1)}$,
$\lambda=\lambda^{(2)}/\lambda^{(1)}$, $a=h^{(2)}/h^{(1)}$, $\rho=\rho^{(2)}/\rho^{(1)}$,
$\mu=\rho^{(2)}\nu^{(2)}/\rho^{(1)}\nu^{(1)}=\rho \nu$. The Prandtl number is
defined with respect to the liquid $1$, i.e. $Pr= \nu^{(1)}/\kappa^{(1)}$, as
well as the Rayleigh number $Ra=\frac{\alpha^{(1)}\beta g (h^{(1)})^4}{\nu^{(1)}\kappa^{(1)}}$ and the Marangoni number $Ma= -\frac{d \sigma}{d T}
\frac{\beta (h^{(1)})^{2}}{\rho^{(1)}\nu^{(1)}\kappa^{(1)}}$.

\section{The basic state}
\label{sec.basicstate}

When a horizontal gradient is imposed, a stationary basic flow sets in each liquid \cite{SmithDavis} 
with a horizontal velocity component depending on the vertical
coordinate $\ub^{(i)}=(u_{0}^{(i)}(z),0,0)$. On the other hand,
the basic temperature profile is the superposition of the imposed
horizontal gradient and the vertical profile $\tau^{(i)}(z)$
generated by the fluid motion: $T^{(i)}=T_- +x +\tau^{(i)}(z)$ (in
nondimensional form). To calculate $u_{0}^{(i)}(z)$ and
$\tau^{(i)}(z)$ we replace

\begin{eqnarray}
\ub^{(i)}=(u_{0}^{(i)}(z),0,0)\, , \hspace{1cm} T^{(i)}-T_-=x+\tau^{(i)}(z), \label{eq:ut}
\end{eqnarray}

\noindent in (\ref{eq:ndbi1}) and (\ref{eq:ndbi2}) and eliminating the pressure one obtains
the following equations governing the behavior of the velocity and temperature fields in both
liquids:

\begin{subequations}
\begin{eqnarray}
&&\partial_{z^3}u_{0}^{(1)}=Ra\; ,\hspace{1cm}\partial_{z^2}\tau^{(1)}=u_{0}^{(1)};\\
&&\nu\partial_{z^3}u_{0}^{(2)}=\alpha Ra\; ,\hspace{1cm}\kappa \partial_{z^2}\tau^{(2)}=u_{0}^{(2)},
\end{eqnarray}
\end{subequations}

\noindent with the  corresponding boundary conditions

\begin{subequations}
\begin{eqnarray}
&&z=-1\to\;\;\;\; u_{0}^{(1)}=0,\;\;\;\;\tau^{(1)}=0;  \\
&&z=a\to\;\;\;\; u_{0}^{(2)}=0,\;\;\;\;\tau^{(2)}=0; \\
&&z=0\to\;\;\;\; \tau^{(1)}=\tau^{(2)},\;\;\;\;u_{0}^{(1)}=u_{0}^{(2)}, \\
&&\;\;\;\;\;\;\;\;\;\;\;\;\;\;\partial_{z}\tau^{(1)}=\lambda\partial_{z}\tau^{(2)},\;
 \mu\partial_{z}u_{0}^{(2)}-\partial_{z}u_{0}^{(1)}=Ma. \nonumber
\end{eqnarray}
\end{subequations}

\noindent The return flow condition \cite{SmithDavis} which requires  that the net flow through a vertical plane must vanish  in each layer, i.e.


\begin{eqnarray}
\int_{-1}^{0}u_{0}^{(1)}=0\;\;\; , \hspace{1cm} \int_0^{a}u_{0}^{(2)}=0,
\end{eqnarray}

\noi provides two additional conditions that allows us to  calculate explicitly the velocity
and temperature fields $u_{0}^{(i)}$ and $\tau^{(i)}$, whose expressions in terms of $z$ are
found in the Appendix.

  As shown by Eqs. (\ref{eq:u11}-\ref{eq:t22}) the basic velocity  and temperature profiles
are the sum of a thermocapillary term proportional to the Marangoni number $Ma$ and a
buoyancy  term proportional to the Rayleigh number $Ra$.  The velocity
profile is a second order polynomial in $z$ in the interfacial term and of third order in the
buoyancy. Whereas the temperature profile is fourth order in the interfacial term and fifth in
the buoyancy.

 The velocity and temperature profiles depend on the ratios of the various transport
coefficients and depths. For a given experimental set-up the heating does not affect
the shape of the basic velocity and temperature profiles since the temperature gradient
 appears as a constant factor in $u_{0}^{(i)}$
and $\tau^{(i)}$ through the Rayleigh and Marangoni numbers.

 Each velocity profile has a root located at its corresponding rigid wall (as imposed by the
boundary conditions (\ref{eq:bcin-a}-\ref{eq:bcin-b})).
 There is also a second root inside each layer due to the  return flow condition. In absence of
 gravity,
this root is  located at $z=-1/3$ in the lower layer
and at $z=a/3$ in the upper fluid, i.e. one third of the depth counted from the interface, and
the  flow
consists of one convective cell in each layer.  When gravity is taken into account the
position of these roots
depend on the properties of the fluids and a third root can be present, allowing for  a
second convective cell in each layer.
The temperature profiles have also a root located at the
bounding horizontal walls. In presence of gravity
it is  possible to find until five roots in each layer.

\subsection{Velocity and temperature profiles }

\begin{center}
\begin{figure}[t]
\centering
\includegraphics[width=\textwidth]{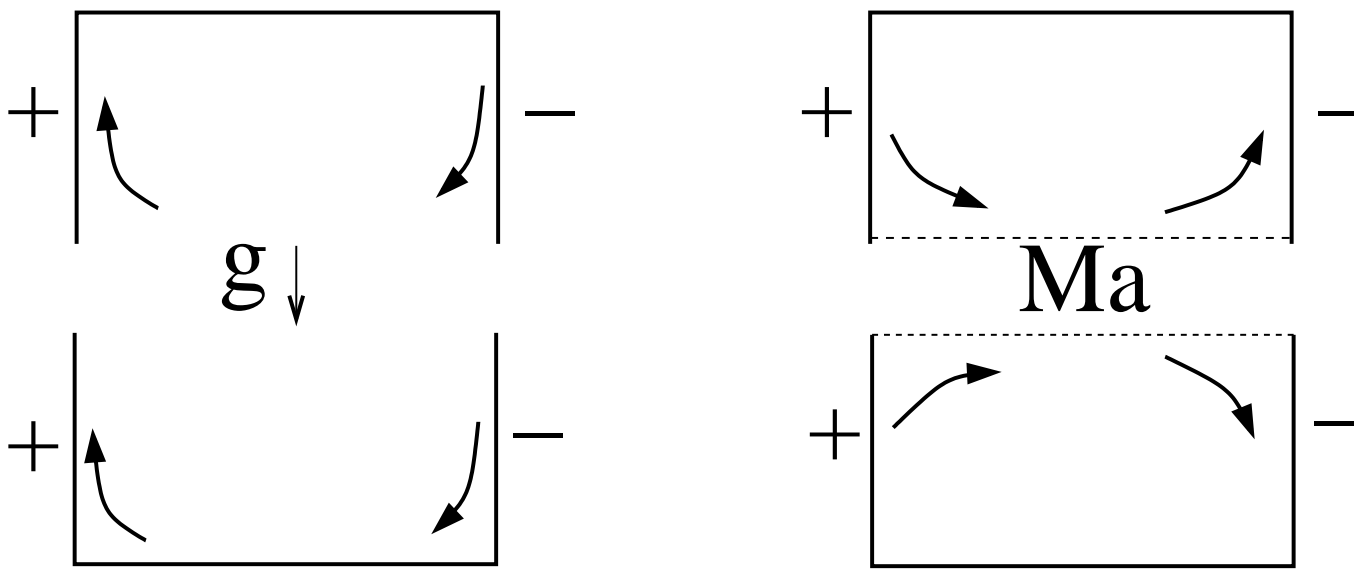} 
\caption{
Driving forces: gravity (left) and interfacial tension (right).
The buoyancy and thermocapillary forces act in the same direction in
  the lower layer, and in opposite direction in the upper layer where they
  are competing.
\label{f:insmech}}
\end{figure}
\end{center}

 The flow is driven by interfacial tension gradients,  and by density differences when the gravity is
acting.  As usually, it is assumed that the surface tension decreases with  temperature ($\gamma > 0$), so that
 a horizontal temperature gradient gives rise to  tangential forces that drive the fluid from hot to cold
regions. Therefore, with only thermocapillary effects present, a general circulation around the interface is established  that
will drag the fluid from the hot to the cold side. In a finite container, by continuity, the fluid in the upper
layer raises near the cold side and falls down near the hot one; on the contrary, in the lower layer it falls
along the cold side and raises up at the hot one. One would therefore observe two counter-rotating cells.
Buoyancy forces also  drive the fluid  from hot (lower density) to  cold regions (higher density).
Since the fluid raises along the hot side and falls at the colder one, in both cells, the buoyancy will favor the
formation of two co-rotating cells.
  In coupled thermocapillary-buoyancy convection, buoyancy and interfacial forces are acting in the same direction in the
lower layer, just like in one-liquid systems with $\gamma > 0$. However they
act in opposite directions in the upper layer, where the forces are competing
(see Fig. \ref{f:insmech}). Thus the liquid 2 exhibits a
 scenario  similar to that of a one-liquid system with $\gamma <  0$ . (This occurs
in some  ceramics and  liquid alloys, like $Ag\, Pb$, \cite{Platten}. )

\begin{center}
\begin{figure}[t]
  \begin{center}
   \subfigure{ 
    \label{fig:u}
      \begin{minipage}[b]{7cm}
      \centering
       \includegraphics[width=7cm]{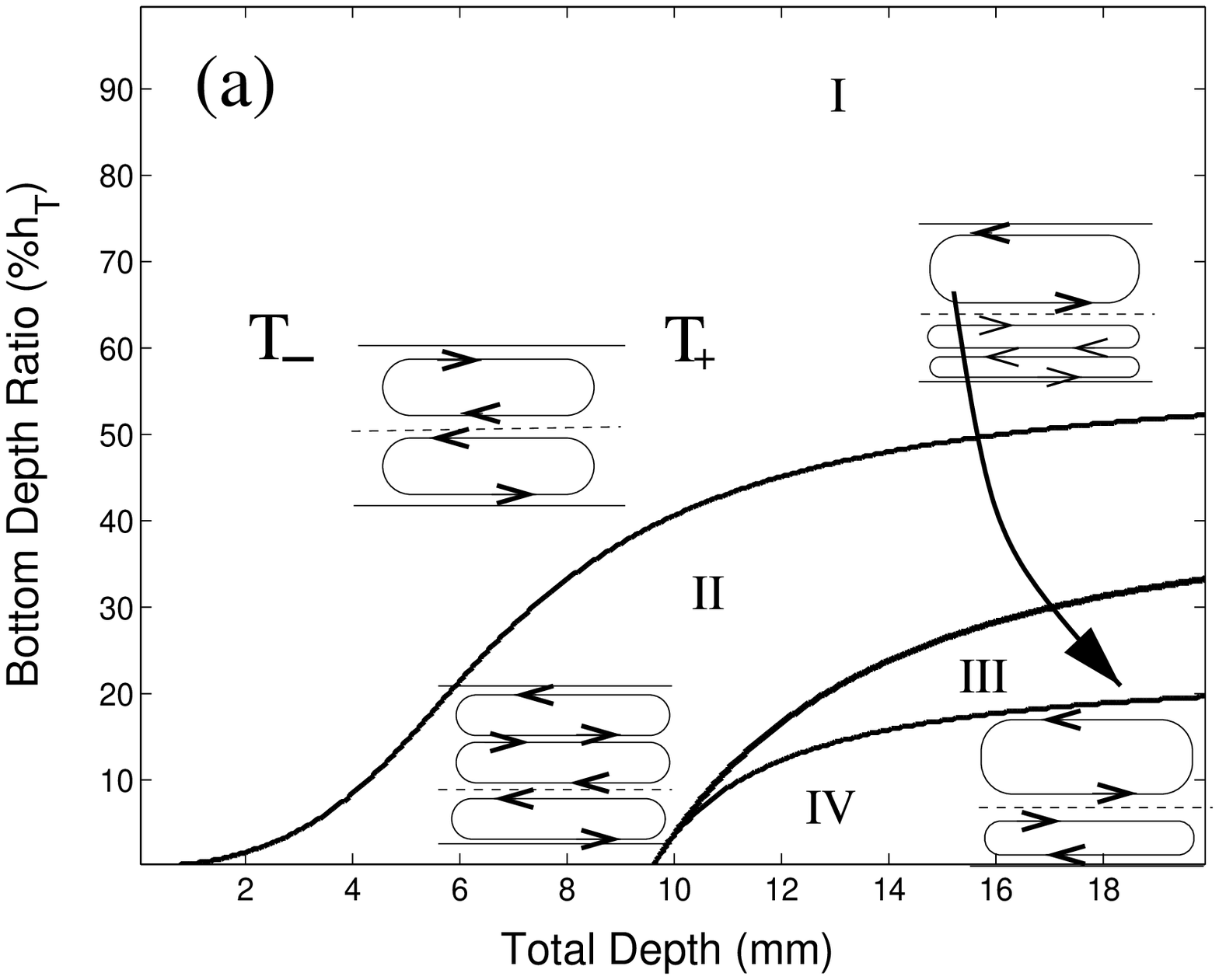}   
     \end{minipage}%
      }
      \subfigure{ 
    \label{fig:t}
      \begin{minipage}[b]{7cm}
      \centering
       \includegraphics[width=7cm]{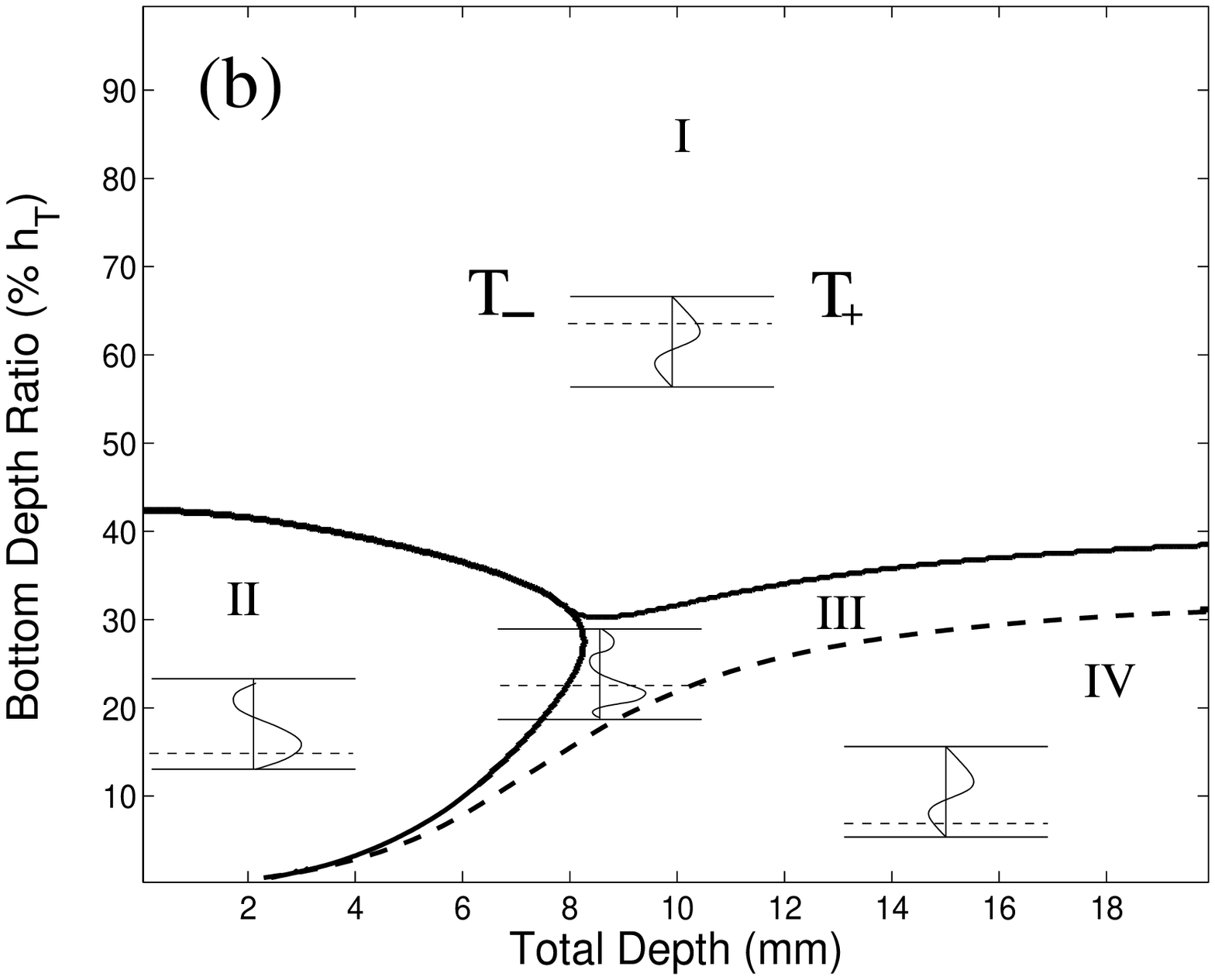}    
     \end{minipage}%
      }
    \end{center}
\caption{ 
Velocity profiles (\ref{fig:u}) and temperature profiles (\ref{fig:t})  of the basic state
  for a system composed by  perfluorinated HT-70 (lower liquid) and $5\;cSt$  silicon oil (upper liquid). The dashed line
gives the position of $M1$ (see the text).
\label{fig:ut} }
\end{figure}
\end{center}

 In Figs. \ref{fig:u} and \ref{fig:t}  are displayed  the basic velocity and temperature profiles.
 Along the horizontal axis is reported the total depth ($h_T$) of the two superposed fluids while the vertical axis
gives the percentage of depth of the lower liquid with respect to the total depth, i.e. the fractional bottom
depth ${\hat h}_1$  (${\hat h}_1 =100/(1+h^{(2)}/h^{(1)})$). The curves refer to a configuration formed by perfluorinated HT-70 (lower fluid) and
$5\;cSt$ silicon oil (upper fluid), this pair of fluids has been used in a recent  experiment
with vertical heating by Juel {\em et al.} \cite{Jue}.    (The physical properties of these  liquids are listed in Tab. \ref{propiedades}.)
 In both figures, four different regions denoted (I) to (IV) are distinguished.  
Let us pay attention to Fig. \ref{fig:u}.
 The region (I) extends to all $h_T$ and covers principally high values of ${\hat h}_1$.
The basic state
 consists of a clockwise convective cell in liquid 2, and a counterclockwise cell in liquid 1;
in this region the direction
of the general circulation is governed by the interfacial forces.
In region (II), the competition between buoyancy and interfacial forces  gives birth  to two
 sublayers in liquid 2. The motion in the sublayer close to the interface is dominated by the
interfacial forces, but
near the top wall it is the buoyancy that contributes to create a second sublayer.
 In regions (I) and (II), the general flow structure in liquid 1 is not changed by the motion of the upper liquid. However for
 high $h_T$ and small ${\hat h}_1$ the motion in layer 1 is driven by the upper liquid.  This is the case
 of regions (III) and (IV), where the direction of circulation close to the interface changes its sign.
 This is so because the buoyancy force acting on liquid 2 is so strong that it does not only drive
  the motion in
 the upper layer but it also modifies the flow direction  near the interface. In (III) two convection cells are found
in liquid 1, but the one close to the lower surface is no longer driven by liquid 2. In region (IV) one observes the
opposite scenario to (I), i.e. two convection cells but with the liquid flowing from
the  cold to  the hot side near the interface. (The behavior of the lower liquid is analogous to that of a single
liquid with $\gamma < 0$ \cite{Platten}.)
 Contrary to the zero gravity problem \cite{DoiKoster} states with one of the two layers  at
rest are not found in the present problem.

Let us now examine the temperature field in the basic state (Fig. \ref{fig:t}).
 The four regions do not coincide with  these of the velocity field
 but they are related. Region (I) covers more than half
 the surface of  Fig. \ref{fig:t}.
  Liquid 2 is unstably stratified (temperature increases with depth), in liquid 1
one finds a lower layer close to the wall which is unstably stratified,
 and above it one distinguishes  a stably stratified sublayer (temperature decreases with depth).
 Region (II) is the only closed area: the lower liquid 1 is stably
 stratified whereas the upper liquid 2 is characterized by  two stable stratified sublayers adjacent to
the wall and the interface,
 in the middle there is an  unstable stratified region, giving raise to a S-shaped profile. Region (III)
 shows the most complicated temperature profile: liquid 2 exhibits two unstably stratified sublayers at
the boundaries and a stable sublayer in the intermediate region, liquid 1
 consists in three sublayers, the intermediate being  stably stratified while the two others
 are unstably stratified. Zone (IV) covers the low ${\hat h}_1$ and almost all $h_T$ values, the global
profile is similar to that of region (I), with the interface located at smaller ${\hat h}_1$ so
 that liquid 1 is unstably stratified.

\section{Evolution equations for the perturbations.} \label{sec.evolequation}

As soon as a horizontal temperature difference $\Delta T$ is applied, convective cells typical of the
basic state set in, but
a further increase of $\Delta T $ may destabilize this basic flow. To analyze the stability of
the reference state with respect to infinitesimally small perturbations, let us write the general solution of the problem
under  the form $\ub^{(i)}=(u_{0}^{(i)}+u^{'(i)},v^{'(i)},w^{'(i)})$, $T^{(i)}-T_{-}=x+\tau^{(i)}+\theta^{'(i)}$,
$p^{(i)}=p_{0}^{(i)}+p^{'(i)}$  where $(u^{'(i)},v^{'(i)},w^{'(i)})$, $\theta^{'(i)}$ and $p^{'(i)}$ denote the
perturbations of velocity, temperature and pressure fields respectively.

 The perturbations are decomposed into a sum
of normal modes $(\ub',\theta',p')=(\ub'(z),\theta'(z),p'(z))\exp{i({\bf  k}\cdot{\bf x}+\omega t)}$,
 where  $\omega$
denotes a  complex frequency and $\bf{k}$ the wavenumber, with component   $k_x$ in the
streamwise
direction and $k_y$ in the spanwise direction. The primes will be dropped for clarity in the
ensuing equations.
 After eliminating the pressure in the momentum equation and omitting non
linear terms, we are left with the following equations for the perturbations,
 \noindent for the lower liquid, 1:
\begin{subequations}
\begin{eqnarray}
\partial_{z^2}u^{(1)}&=&L^{(1)}\,u^{(1)}+\frac{k_{y}^2}{k^2} Pr^{-1}\partial_{z}{u_{0}^{(1)}}w^{(1)}+i\,\frac{k_{x}}{k^2}\left(\partial_{z^3} w^{(1)} -L^{(1)}\,\partial_{z}\,w^{(1)}\right),\\
\partial_{z^4}w^{(1)}&=&(L^{(1)}+k^2)\partial_{z^2}w^{(1)}-L^{(1)} k^2 w^{(1)} +Ra\,k^2 \theta^{(1)}-i\,k_{x} Pr^{-1}w^{(1)}\partial_{z^2}{u_{0}^{(1)}}, \\
\partial_{z^2}\theta^{(1)}&=&(k^2+i\,(\omega+u_{0}^{(1)}k_{x}))\theta^{(1)}+u^{(1)}+\partial_{z}{\tau^{(1)}}w^{(1)},
\end{eqnarray}
\end{subequations}
\noindent for the upper liquid, 2:

\begin{subequations}
\begin{eqnarray}
\nu\partial_{z^2}u^{(2)}&=& L^{(2)} u^{(2)}+\frac{k_{y}^2}{k^2}Pr^{-1}\partial_{z}{u_{0}^{(2)}}w^{(2)}+i\,\frac{k_{x}}{k^2}\left(\nu\partial_{z^3} w^{(2)} -L^{(2)}\,\partial_{z}\,w^{(2)}\right),\\
\nu\partial_{z^4}w^{(2)}&=&(\nu k^2+L^{(2)})\partial_{z^2}\,w^{(2)}-k^2 L^{(2)} w^{(2)}+Ra\,k^2\alpha\,\theta^{(2)}  -i\,k_{x} Pr^{-1}w^{(2)}\partial_{z^2}{u_{0}^{(2)}},\\
\kappa\partial_{z^2}\theta^{(2)}&=&(\kappa
k^2+i\,(\omega+u_{0}^{(2)}k_{x}))\theta^{(2)}+u^{(2)}+\partial_{z}{\tau^{(2)}}w^{(2)},
\end{eqnarray}
\end{subequations}

\noindent wherein  the following notation has been used
\begin{eqnarray*}
L^{(1)}&=&k^2+i\,Pr^{-1}(\omega+u_{0}^{(1)}k_{x}),\\
L^{(2)}&=&\nu k^2+i\,Pr^{-1}(\omega+u_{0}^{(2)}k_{x}).
\end{eqnarray*}

\noindent The corresponding boundary conditions are

\begin{subequations}
\begin{eqnarray}
&&z=-1\to\;\;\;\, u^{(1)}=w^{(1)}=\partial_z w^{(1)}=\theta^{(1)}=0; \\
&&z=a\to\;\;\;\;\;\;  u^{(2)}=w^{(2)}=\partial_z w^{(2)}=\theta^{(2)}=0; \\
&&z=0\to\;\;\;\;\;\;u^{(1)}=u^{(2)},\;\;\;\partial_z w^{(1)}=\partial_z
w^{(2)},\;\;w^{(1)}=w^{(2)}=0, \\
&&\hspace{1.8cm} \theta^{(1)}=\theta^{(2)},\;\partial_{z}\theta^{(1)}=\lambda \partial_{z}\theta^{(2)},\; \mu
\partial_{z}{u^{(2)}}-\partial_{z}{u^{(1)}}=i k_{x} Ma\theta^{(1)},\;\;\;\; \nonumber \\
&&\hspace{1.8cm} \mu\partial_z^2{w^{(2)}}-\partial_z^2{w^{(1)}}=Ma \,k^2
\theta^{(1)}. \nonumber
\end{eqnarray}
\end{subequations}


We are faced with an eigenvalue  problem which is solved by means of a tau spectral method, by approximating
the eigenfunctions with Chebyshev polynomials  of order 16. The marginal curves are found by searching the
set of values $k_x,k_y,Ma$ for which the rate of temporal growth is zero.

\section{Results of the linear analysis}

Three mechanisms are able to destabilize the system under study.
Basic temperature profiles in Fig. \ref{fig:t}  present unstably stratified zones (temperature
increases with depth) and other zones which are stably stratified (temperature decreases with depth),  
in  the unstably stratified regions the  Rayleigh-B\'enard instability may arise.
 A second mechanism is the usual  B\'enard-Marangoni instability when the interface is colder than the adjacent
fluid.
 Finally, a third destabilizing mechanism arises even when the core of the layers 
is colder than the interface,  provided the velocity flow
 is strong enough to overcome the stabilizing vertical temperature field
\cite{Smith1}.

Owing to the  great number of
parameters involved in a two-layer configuration, the results will be discussed in the case
of selected pairs of liquids used in experiments. The  calculations have been made for the pair
    $5\; cSt$  silicon oil (upper liquid) and perfluorinated
HT-70 (lower liquid),  but later on we also consider the combinations:
water (upper fluid) with  perfluorinated hydrocarbon Fc-75 (lower liquid), and
  n-hexane (upper liquid) with  acetonitrile (lower liquid).
   The parameter values of these fluids are gathered in Tab. \ref{propiedades}. 
 Deformability is generally quantified 
by means of the so-called crispation number, defined as $Cr=\frac{\rho^{(1)} \nu^{(1)} \kappa^{(1)}}{\sigma_0 h^{(1)}}$. As shown in Tab. \ref{propiedades} the values of $Cr$ are very small for the pairs of liquids
examined in the present work. Then the hypothesis of nondeformability of the interface is reasonable as it has
 been shown en earlier publications \cite{Carneiro,RegnierDauby}.
 Moreover,  we will not pay attention to the particular
 ${\hat h}_1$ values close to $0\%$ or $100\%$ because in these cases thin film motions
 (outside the scope of the present analysis) would appear.

\subsection{Stationary longitudinal rolls}

 First, we consider 2-D perturbations characterized by $k_x =0$ and $\omega=0$
(longitudinal stationary rolls). Our analysis extends for two layer systems that of
Gershuni {\em et al.} \cite{Gershuni}, who considered a one-single fluid
layer bounded by horizontal rigid-free and perfectly thermal conductive surfaces,
and Mercier and Normand \cite{Mercier}, who considered the heat
transfer at the free surface by means of a Biot number.

\begin{center}
\begin{figure}[t]
  \begin{center}
   \subfigure{ 
    \label{fig:2dMa}
      \begin{minipage}[b]{6.25cm}
      \centering
       \includegraphics[width=6.25cm]{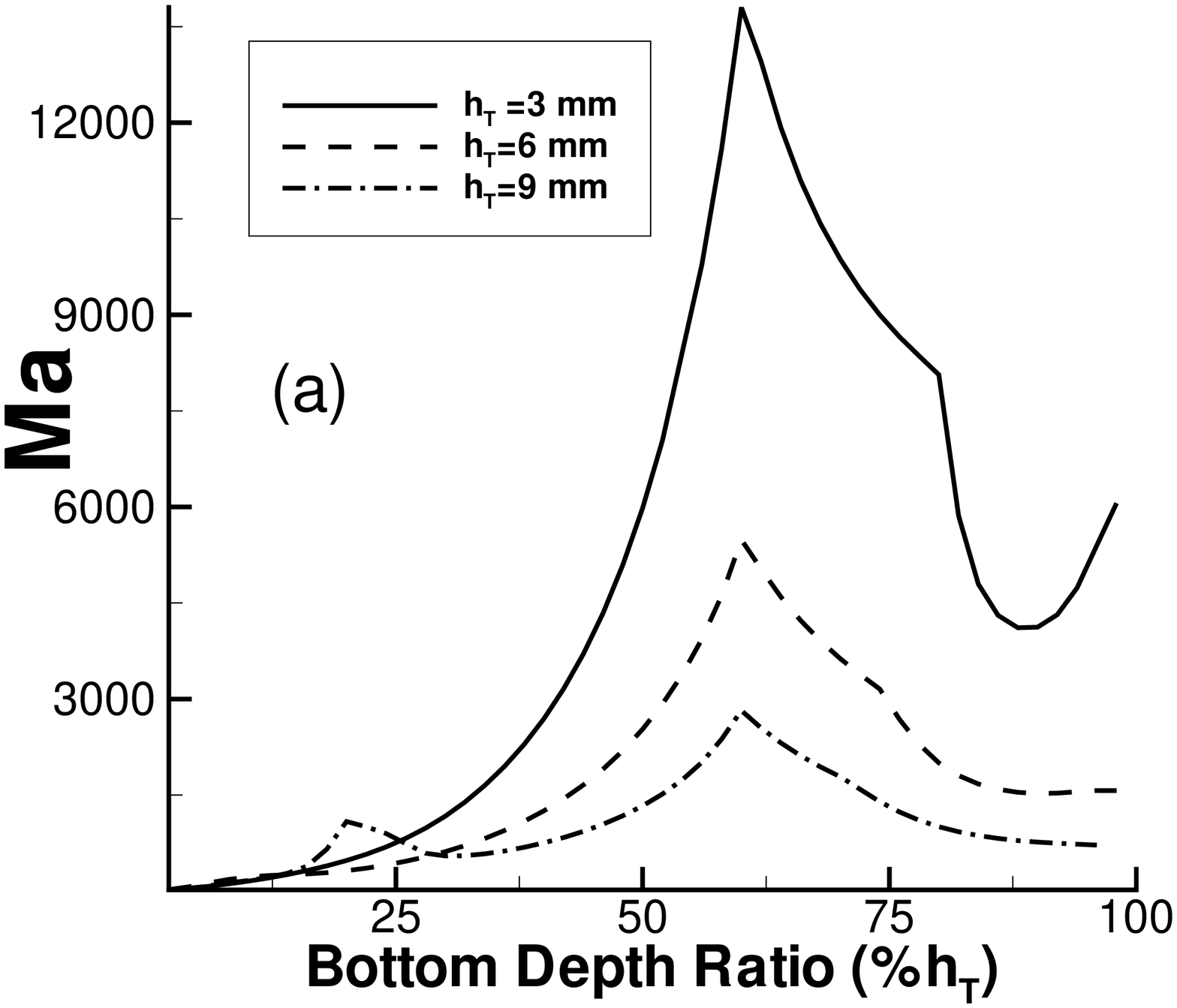}
     \end{minipage}%
      }
      \subfigure{ 
    \label{fig:2dky}
      \begin{minipage}[b]{6.25cm}
      \centering
       \includegraphics[width=6.25cm]{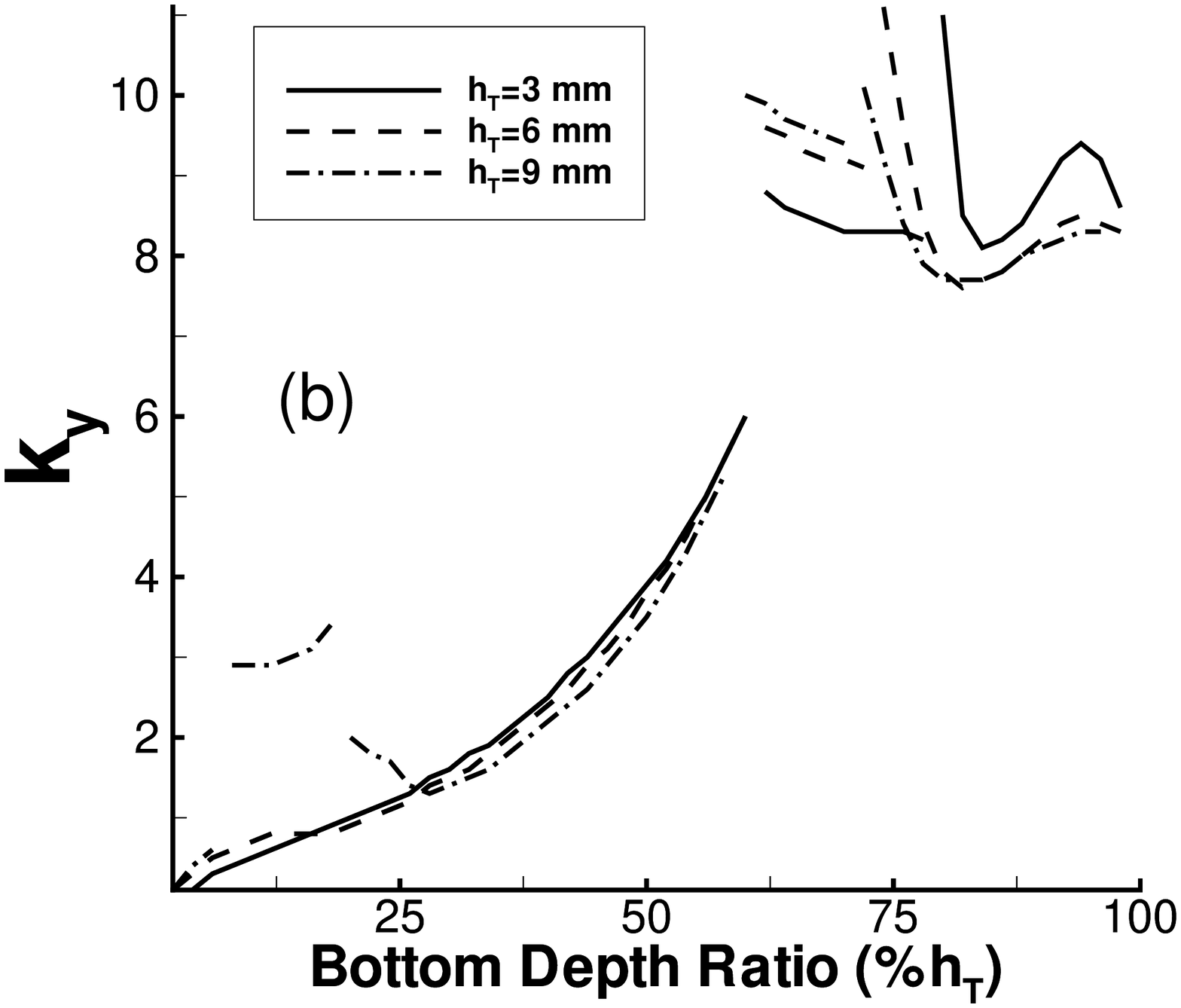}    
     \end{minipage}%
      }
    \end{center}
\caption{ 
Critical Marangoni (left) and critical wavenumber (right)
 for 2-D  longitudinal  perturbations ($k_x=0$)  as a function of the bottom depth
 ratio, for  three  different values of the total depth. The results correspond to the pair HT-70 (lower liquid)
and $5\; cSt$  silicon oil (upper liquid).
\label{fig:2d}}
\end{figure}
\end{center}

 Fig. \ref{fig:2d} displays the critical Marangoni and the critical
spanwise wavenumber  versus the fractional bottom depth for several values of the total
 depth $h_T$ ($3,6$ and $9\; mm$).
   For total depths of $3$ and $6\;mm$, one observes three stationary branches,
while for  $h_T=9\;mm$   an additional  branch  is clearly seen for low  ${\hat h}_1$.
 Fig. \ref{fig:2dMa}
shows that the critical Marangoni number decreases with $h_T$. This is due to the
destabilizing effect of the increase of depth in Rayleigh-B\'enard instability.  All the curves exhibit the most  stable zone
for  ${\hat h}_1\approx 60\%$.
The jump from one branch to another is better appreciated  by the discontinuities in $k_y$ in Fig. \ref{fig:2dky}.
The spanwise wavenumber  $k_y$ for $h_T=3\;mm$ and $h_T=6\;  mm$
  increases with ${\hat h}_1$ in the first branch, however it
decreases in the second one together with that of $h_T=9\;mm$, the dependence with ${\hat h}_1$ in third 
branch is more involved for the three depths.
In the branch jump the marginal Marangoni curve is bimodal
\cite{Mercier}, switching the absolute minimum between two
wavenumbers.  This modal change gives rise to different states in
the two-layer system.

\begin{center}
\begin{figure}[t]
  \begin{center}
   \subfigure{ 
    \label{fig:2d14}
      \begin{minipage}[b]{6.25cm}
      \centering
\includegraphics[width=6.25cm]{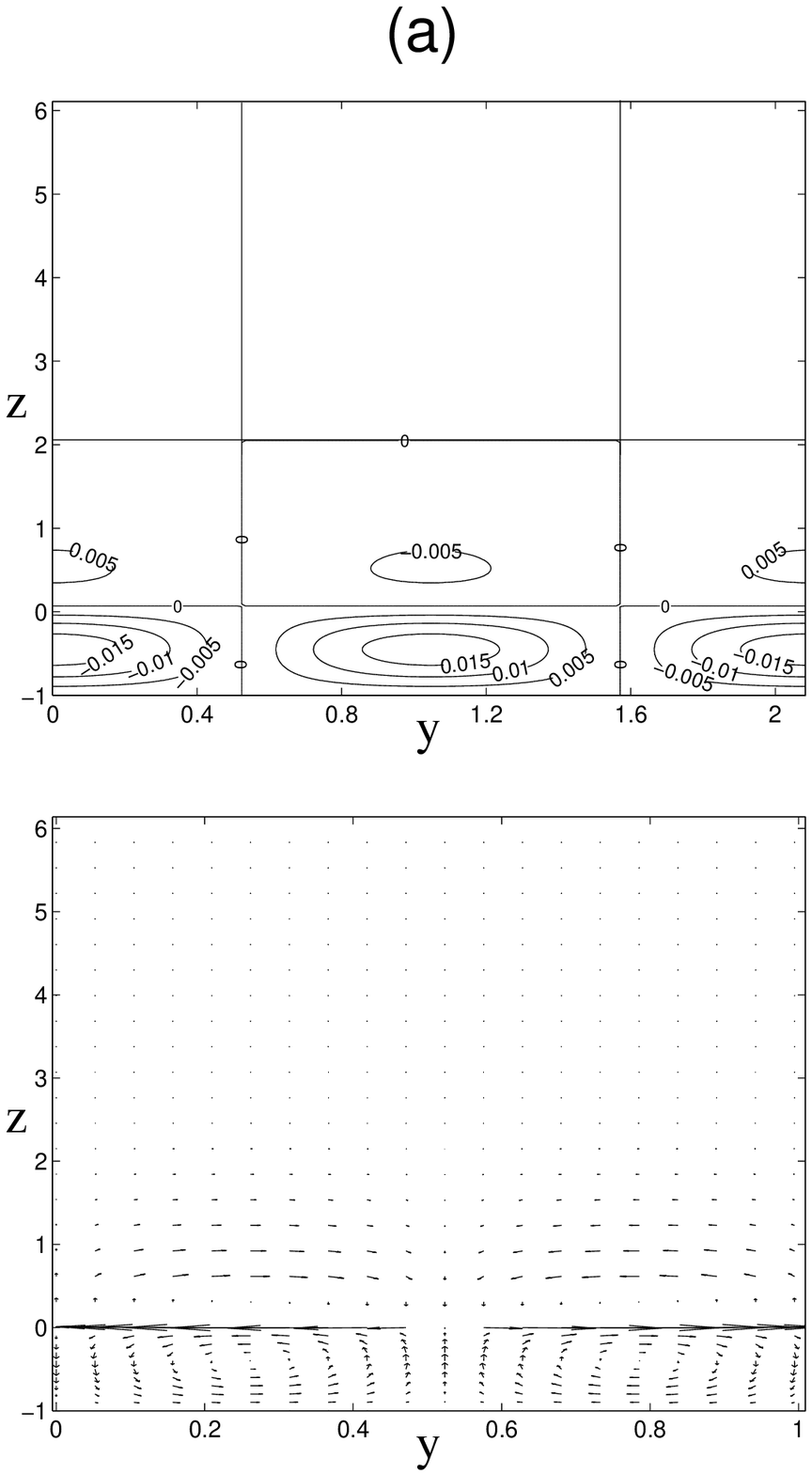}    
     \end{minipage}%
      }
      \subfigure{ 
    \label{fig:2d80}
      \begin{minipage}[b]{6.25cm}
      \centering
\includegraphics[width=6.25cm]{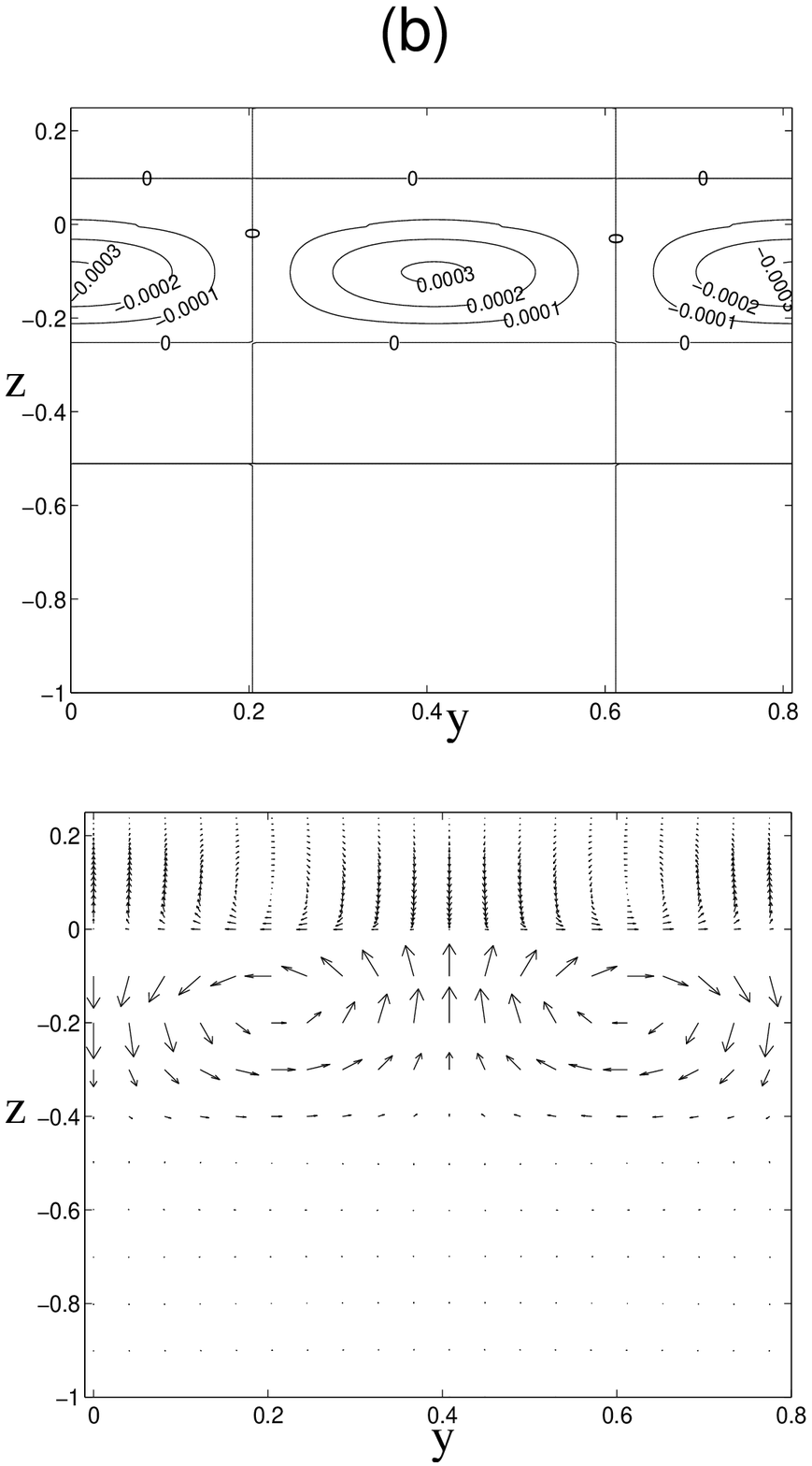}    
     \end{minipage}%
      }
    \end{center}
\caption{
 Isotherms (above) and velocity fields (below) in
the plane z-y are represented  at critical thresholds for
 $5\;cSt$  silicon oil/HT-70 system and  $h_T=9\;mm$, for two bottom depth ratios corresponding to different branches:   ${\hat
h}_1=14\%$ (right ), ${\hat h}_1=80\%$ (left).
\label{fig:2dper}}
\end{figure}
\end{center}

 In Fig. \ref{fig:2dper} are represented the isotherms and the velocity
fields in the plane $z-y$ as observed at the critical threshold. At small fractional bottom
depths
( ${\hat h}_1 =14\%$ in Fig. \ref{fig:2d14}) strong
temperature gradients are found in the bulk of liquid 1, whereas in liquid 2 they are
concentrated near the interface $z=0$,
where they are less important. A similar behavior is exhibited by the velocity fields.
For large fractional bottom depths ( ${\hat h}_1 =80\%$ in Fig. \ref{fig:2d80}) located at other branch
the temperature gradients are concentrated
in  liquid 1 near the interface and do not penetrate into the bulk of the lower liquid, the velocity
field is the largest in the lower liquid and is felt of roughly the same depth than the temperature
gradients. Other stationary branches exhibit different kind of behaviors.

\subsection{Three dimensional perturbations}

\begin{center}
\begin{figure}[t]
\centering
\includegraphics[width=10cm]{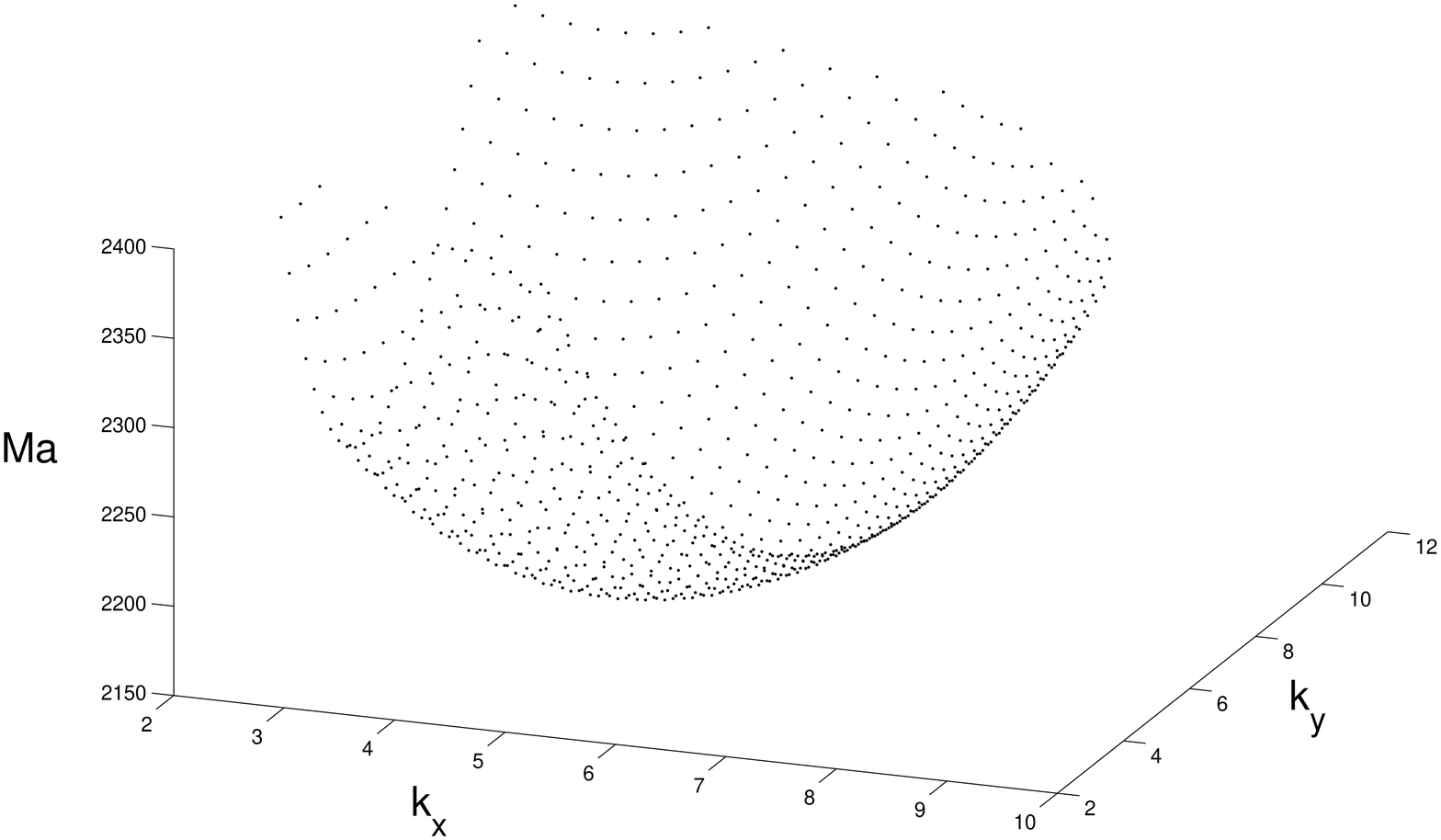}   
\caption{
 Surface of neutral stability for three dimensional perturbations.
In this example $h_T =6\;mm$, and ${\hat h}_1=78\%$. The breaking of $k_x$, $k_y$ symmetry
is clearly appreciated. The  stability threshold is located at
${\bf k}=(4.2,7)$.
\label{fig:marginal}}
\end{figure}
\end{center}

However, one cannot exclude the possibility that the most unstable disturbances are three dimensional.
 If follows that  a minimization process with respect to $k_x$ and $k_y$ is required to find the
critical parameters. A typical stability surface is shown in Fig. \ref{fig:marginal}. In presence of
lateral heating the
reflection symmetry in the direction $x$ is broken. Hence the marginal surface is only symmetrical
under the changes  $k_y\rightarrow -k_y$. The marginal surface is, in general, more complicated   because
several intermingled branches arise. The position of the minimum is given by the $k_x$ and $k_y$ components
 of the critical wavenumber, which in the  case of oscillatory
instability also determine the  direction
of propagation $\varphi$ of the waves.
Recalling that the positive $x$ axis is directed from the cold to hot side according to
Eq. (\ref{eq:ut}) the hydrothermal waves will propagate from  cold to hot  for
$\varphi \in [0^{o},90^{o}]$, in contrast for $\varphi\in [90^{o},180^{o}]$ the direction of
propagation is the opposite. Note that the symmetry $(y\rightarrow -y)$ of the linear problem makes that the
waves with wavenumber $(k_x,\pm k_y)$ become unstable simultaneously. From now on, we will restrict to $\varphi
\in [0^{o},180^{o}]$ without lost of generality.

\begin{center}
\begin{figure}[!t]
  \begin{center}
   \subfigure{
    \label{fig:3d9mm}
      \begin{minipage}[b]{\textwidth} \label{fig:Maht9}
      \centering
     \includegraphics[width=1\textwidth]{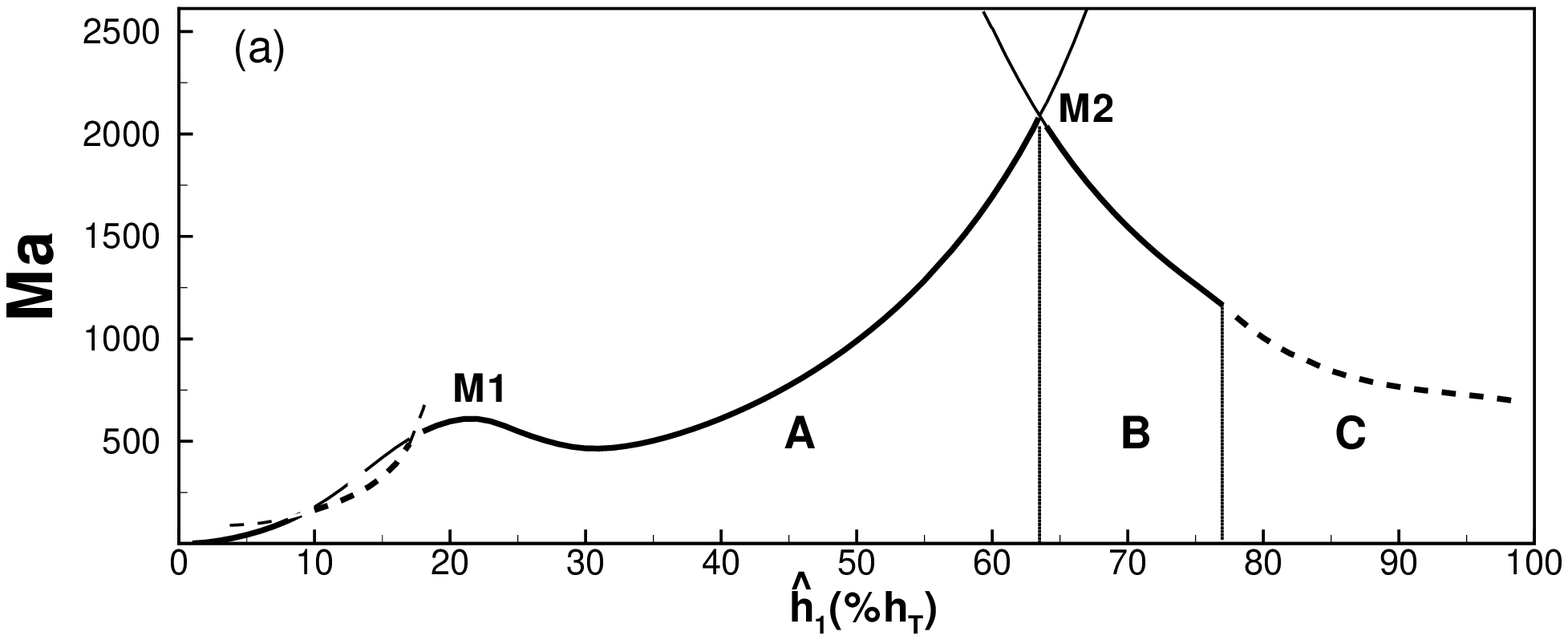}  
     \end{minipage}%
      } \\
      \subfigure{
      \begin{minipage}[b]{\textwidth}
      \centering
       \includegraphics[width=1\textwidth]{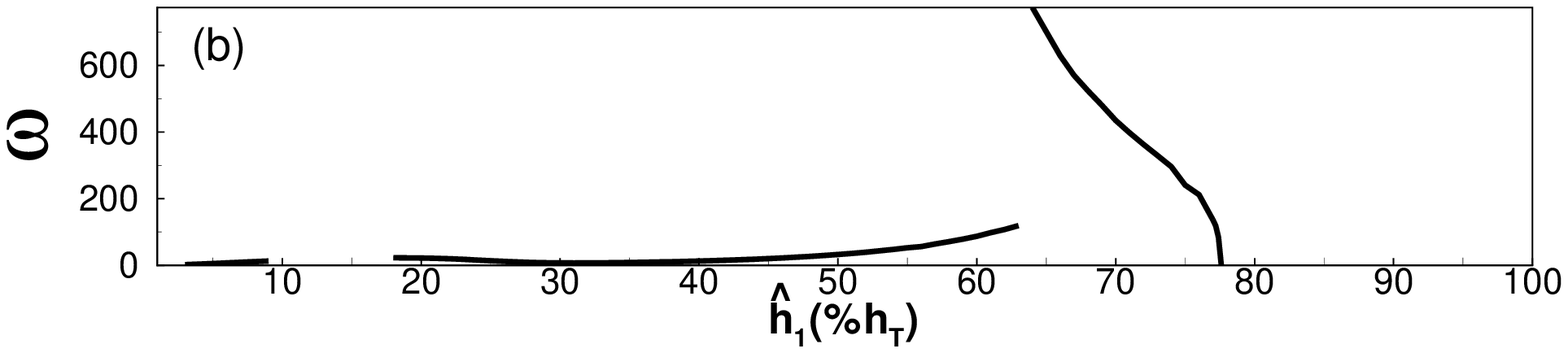}
     \end{minipage}%
      }\\
      \subfigure{
      \begin{minipage}[b]{\textwidth}
      \centering
       \includegraphics[width=1\textwidth]{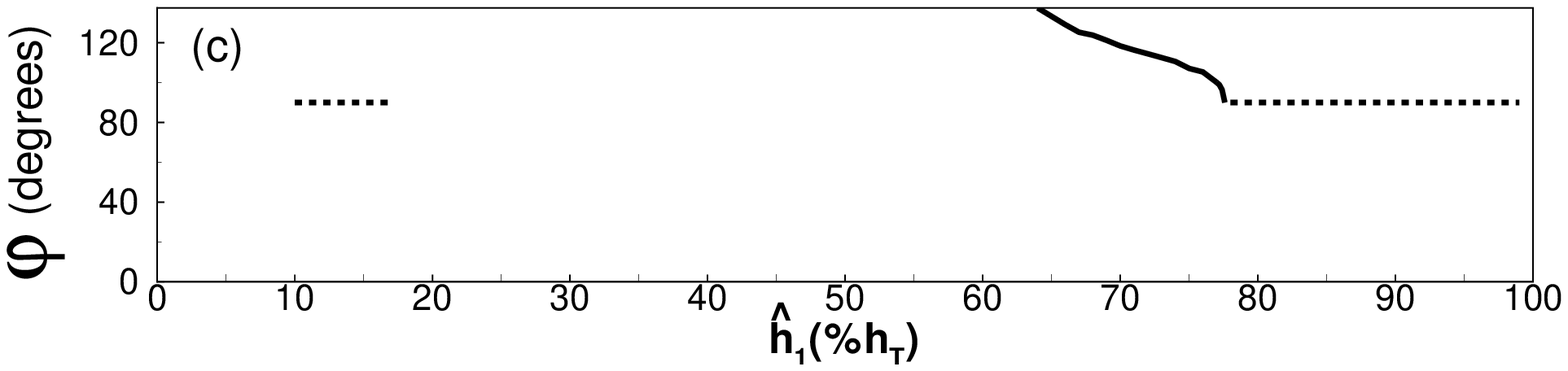} 
     \end{minipage}%
      }\\
    \subfigure{
      \begin{minipage}[b]{\textwidth}\label{fig:modk}
      \centering
       \includegraphics[width=1\textwidth]{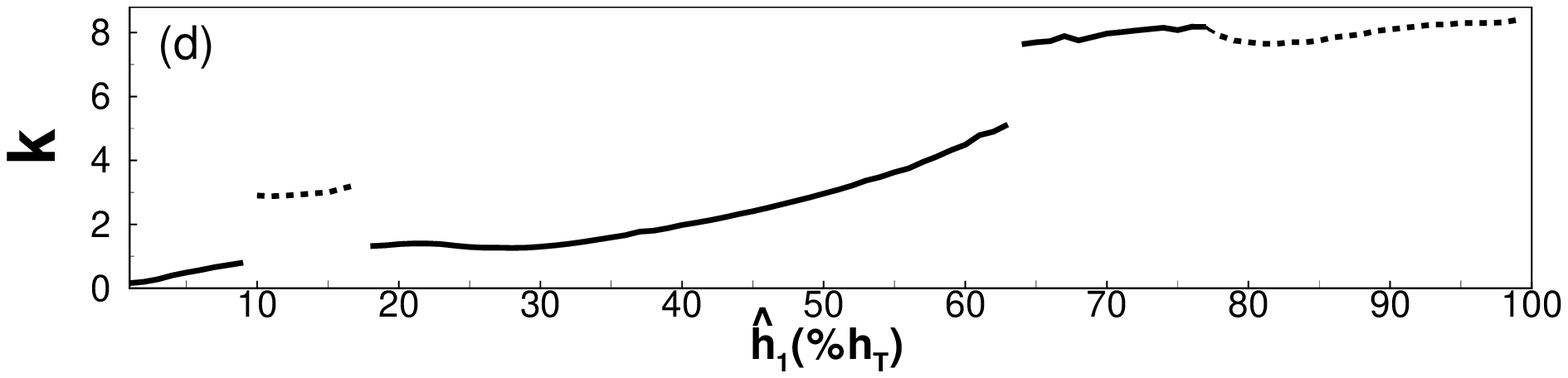}
     \end{minipage}%
      }
    \end{center}
\caption{
Critical parameters for the system $5\;cSt$
silicon oil (upper liquid) and
  HT-70 (lower liquid) with total depth $9\;mm$. (a) Critical Marangoni, (b)
  critical frequency, (c) angle of the critical wavevector and (d) modulus of
  the critical wavevector. Solid  line: oscillatory behavior, dashed line:
  stationary behavior. In the points where there is an interaction between two branches
  are displayed their extensions in the Marangoni curve. Different characteristic  patterns are obtained in  regions
   A, B and C ( see the text).
\label{fig:9mm}}
\end{figure}
\end{center}

In Fig.  \ref{fig:3d9mm} are represented the critical Marangoni number, frequency, angle and modulus of
the wavenumber for the system HT-70 (lower liquid) and $5\;cSt$ silicon oil (upper liquid)  with
$h_T =9\; mm$.  The critical Marangoni curve exhibits a local maximum  $M1$ and a local minimum at
${\hat h}_1=31\%$. The Marangoni number reaches its maximum value in $M2$, which is a crossing point
between two branches.

Beyond $M2$, the behavior of the critical Marangoni is the same as in a
 one-layer system filled
by the lower fluid. In such a case the critical Marangoni also
decreases  and the spanwise wavenumber increases with the depth \cite{Gershuni}.

\begin{center}
\begin{figure}[t]
  \begin{center}
      \subfigure{  
      \begin{minipage}[b]{8cm}   
      \centering
       \includegraphics[width=8cm]{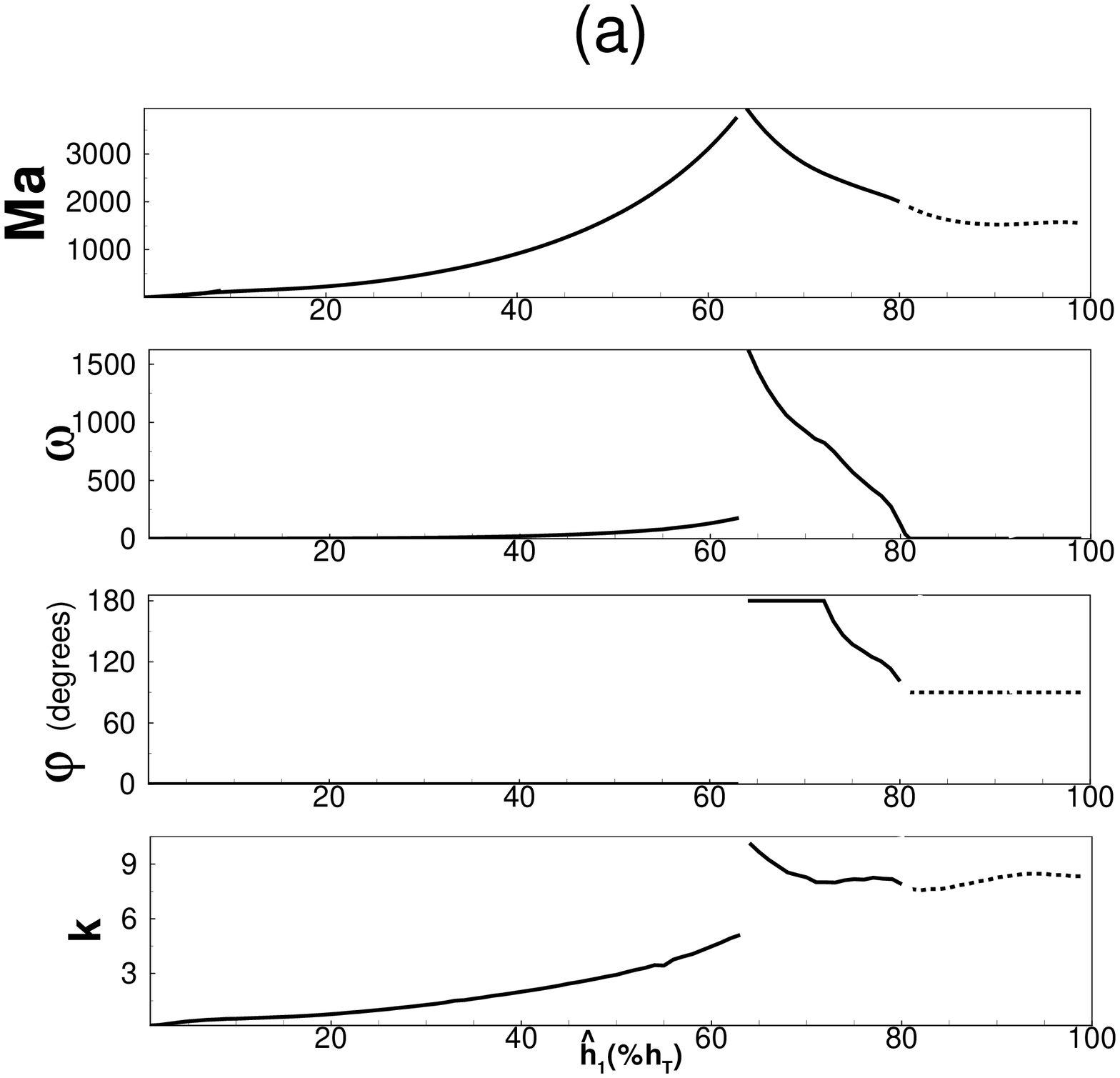} 
     \end{minipage}%
      }
   \subfigure{   
      \begin{minipage}[b]{8cm}   
      \centering
       \includegraphics[width=8cm]{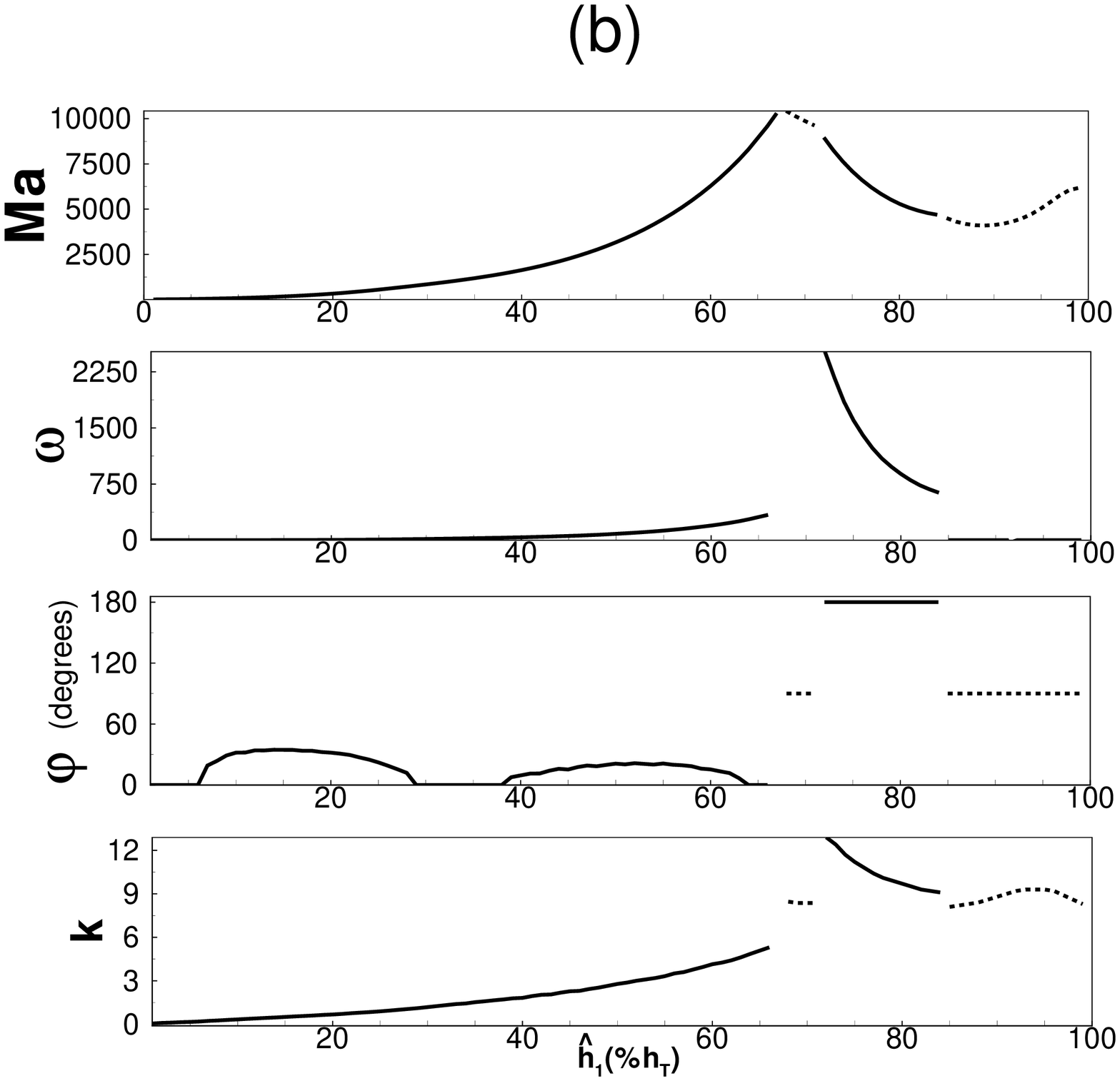}  
     \end{minipage}%
      }
    \end{center}
\caption{
Critical parameters for the system   HT-70 (lower liquid) and $5\; cSt$  silicon oil (upper liquid)
 with total depth $6\;mm$ (a) and $3\;mm$ (b). Solid  line: oscillatory behavior, dashed line:
  stationary behavior.
 \label{fig:3-6mm}}
\end{figure}
\end{center}

The position of $M2$ depends on the physical properties of both fluids.
 To study the effect
of the transport coefficients, i.e. viscosity $\nu$ and thermal diffusivity $\kappa$ on the position
of $M2$ we have considered an ideal
 system formed by two layers of $5\;cSt$   silicon oil which differ only by their values of
$\nu$ and  $\kappa$.  In
Tab. \ref{nu-kappa} are given the positions of $M2$ for different values
of the transport coefficients. By increasing the thermal diffusivity and viscosity of
the lower liquid, the position of $M2$ is shifted towards higher values of ${\hat h}_1$; on the contrary
if $\nu$ and $\kappa$ are increased
in the upper liquid the shift is towards lower values of ${\hat h}_1$. The location of $M2$
depends also on the geometry
of the system: an increase of $h_T$ produces a shift towards a greater ${\hat h}_1$, however for
high $h_T$ values the shift becomes saturated, as shown by the Marangoni curves of
Figs.  \ref{fig:9mm} and \ref{fig:3-6mm}.

  Examination of the vertical profile of temperature of the basic state
allow us to determine the positions of the local maximum $M1$ at $21\%$ and of the local minimum at about
$31\%$. Fig. \ref{fig:t} displays
the vertical profile of temperature; it is seen that, for  $h_T =9\;mm$, starting from the region (I)
and decreasing  ${\hat h}_1$  up to $31\%$ the unstable vertical profile of the upper fluid splits in two
smaller unstable
sublayers and a third one in the middle which is stable. The shape of the temperature profile remains
unchanged in liquid 1.

In the configuration represented by region (III), there is a
greater extension of the stable regions than in (I), thus the
scenario in (III) consists in an increase of $Ma$ due to the progressive stabilization of the upper layer
when ${\hat h}_1$ decreases.
 The
border between regions (I) and (III) gives the position of the local
minimum of Fig. \ref{fig:Maht9}.
  However when region (IV) is attained the stratification in liquid 1 becomes
unstable and the whole system is  more and more unstable, with a decrease of $Ma$. This finds its roots
in the two driving forces (thermocapillary and buoyancy), which are both destabilizing and cooperatives for
the perturbations undergone by
 the lower liquid. As a consequence, the border between regions (III) and (IV) of Fig. \ref{fig:t}  gives the position
of $M1$. Inside (IV) the decrease of $Ma$ when ${\hat h}_1$ is lowered reflects the greater importance of the
destabilizing effect of the interfacial tension. The position of $M1$ is predicted by the vertical
temperature profile of the basic state and it will not be influenced by the Prandtl number.

One distinguishes three main regions A, B, C in the curves of Figs. \ref{fig:9mm} and \ref{fig:3-6mm}
giving the critical parameters as a function of the relative depth of the liquids.

\subsubsection{region A }
The first region A extends from the lowest  value of ${\hat
h}_1$ until $M2$. The critical Marangoni increases with the drop
of the depth of the upper liquid because of the decrease of the
buoyancy effects in it, indicating that the upper liquid dominates
the dynamics in A. The pattern in this region consists of
hydrothermal waves with zero or small angles of propagation. They
propagate from cold to hot regions with an angle parallel to the
gradient of temperature for $h_T=6,9\;mm$, deviating to greater
angles when the thermocapillary effects are relatively more
important, as for $h_T=3\;mm$. Concerning the oscillatory
modes, the angle of propagation is similar to single liquids
with a high $Pr$ \cite{Smith1}.
 As seen in
Fig. \ref{fig:modk}, region A is characterized by an
increase of $|\bf{k}|$ with ${\hat h}_1$, with a small dependence
of $|\bf{k}|$ on $h_T$. At $h_T =9\;mm$, where the buoyancy
effects are more important, one finds a range of depth
ratios $6\% <{\hat h}_1 < 16\%$ where the critical modes take the
form of stationary rolls. At the borders of this window, there is a
codimension-two (cod-2) point formed by the interaction between a Hopf mode and a
stationary mode of different wavenumbers. Such cod-2 points were
also found in two-liquid layers heated from below
\cite{Fujimura}.

\subsubsection{region B }
Region B is characterized by a strong jump of the
frequency  at $M2$ for $h_T =6,9\;mm$, and weakly
shifted by $\Delta {\hat h}_1 =4$  for $h_T =3\;mm$.
For $h_T=6,9\;mm$,  $M2$ is characterized by a pair of Hopf modes with finite
 and different wavenumbers and frequencies which are simultaneously
 critical. At this point of cod-2  the configuration is the most stable
 with the greatest $Ma$. The difference between the wavenumbers of the two Hopf modes is due
to the occurrence of a large spanwise component for $h_T=9\;mm$, while for $h_T=6\;mm$ there exists only a jump in the
streamwise component.
 To our knowledge this kind of interaction between two
oscillatory modes has not yet 
received attention in  B\'enard--Marangoni problems.
 For $h_T =3\;mm$, the presence of  a small stationary branch near
 $M2$ prevents the existence of this kind of interaction, and the cod-2
 point is generated by a stationary mode and a Hopf mode.
 Concerning the hydrothermal waves, the corresponding angles of
 propagation are greater than the angles of propagation in region A.
 The direction of propagation of the hydrothermal
 waves in this region is reversed,  propagating from the hot to the cold side. The  width of region B
 decreases with  $h_T$, from $\Delta {\hat h}_1 =18\%$ for $h_T =9\;mm$ to $\Delta {\hat h}_1 =14\%$
 for $h_T =3\;mm$. The modulus of the wavenumber varies in a more complicated way
 than in the region A, and is more sensitive to the total depth.
 In region B, the strong competition between the two
 layers is responsible for the
oscillatory convection.

\subsubsection{region C}
 The transition from the second region B to the third region C is smooth with a monotonous
decrease of the frequency of the oscillatory  branch.
 The third region is characterized by stationary longitudinal rolls and
extends up to the highest ${\hat h}_1$. It spreads towards lower ${\hat h}_1$
when the total depth is increased. The dynamics of this region is dominated by the lower layer.
  The qualitative behavior of the modulus $k$ of the
wavenumber is practically unaffected by the overall depth.

\begin{center}
\begin{figure}[t]
  \begin{center}
   \subfigure{ 
      \begin{minipage}[b]{8cm}  \label{fig:fc-water} 
      \centering
       \includegraphics[width=8cm]{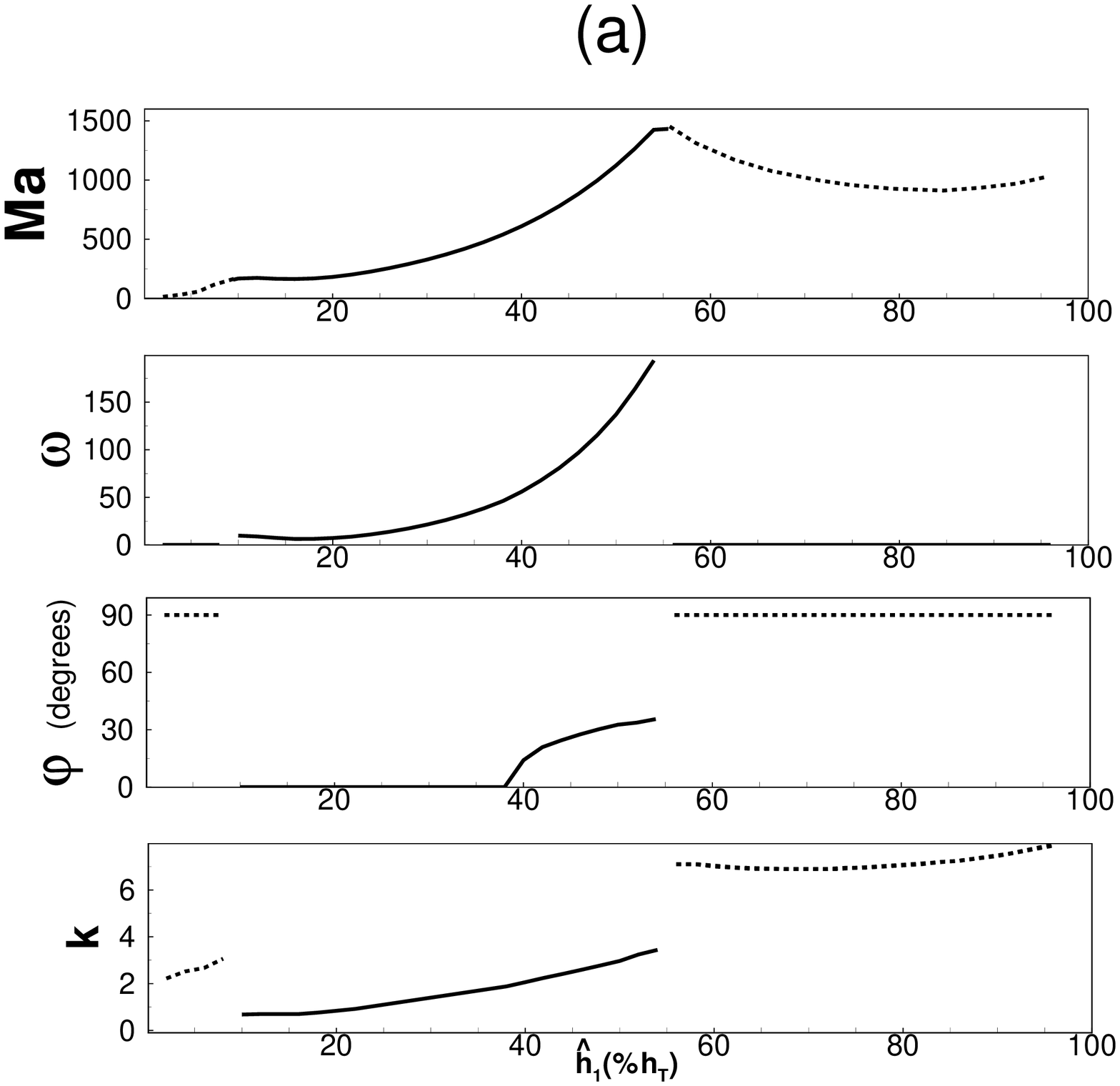}  
     \end{minipage}%
      }
      \subfigure{ 
      \begin{minipage}[b]{8cm} \label{f:ace-hex}    
      \centering
       \includegraphics[width=8cm]{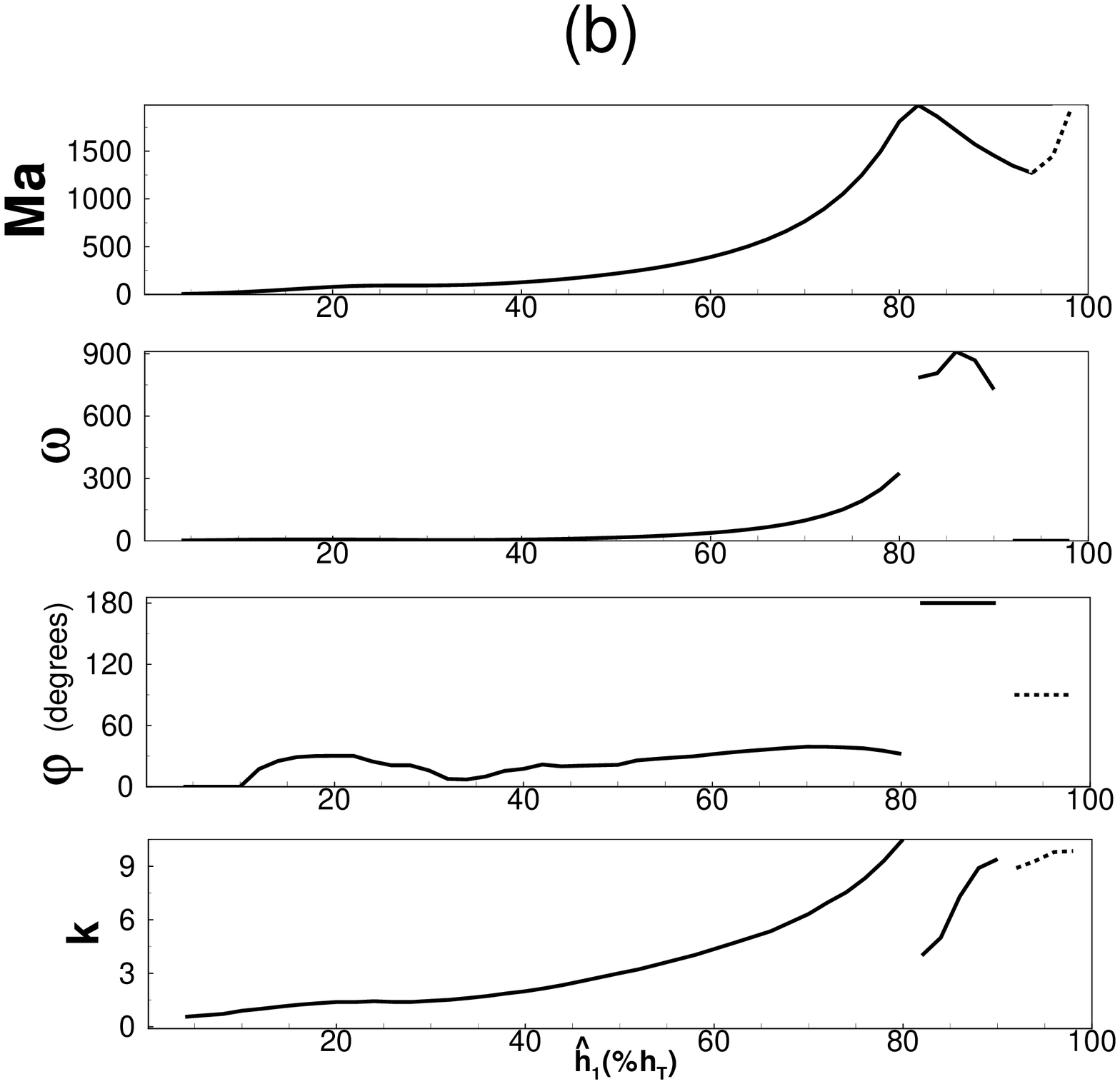}  
     \end{minipage}%
      }
    \end{center}

\caption{
Critical parameters for the pair Fc-75 (lower liquid) and water (upper
  liquid) (a) and for the pair acetonitrile (lower liquid) and
      n-hexane (upper liquid) (b). The total depth for both configurations is $h_T=6\;mm$.
}
\end{figure}
\end{center}

In Fig. \ref{fig:fc-water} are shown the critical parameters for the bilayer formed by
 perfluorinated hydrocarbon Fc-75 (lower liquid) and water (upper fluid),  whereas in Fig. \ref{f:ace-hex}
are reproduced
the results for the couple acetonitrile (lower liquid) and n-hexane (upper liquid). Both setups
have been experimentally studied in the case of a vertical heating \cite{And,Tok,Jue}. The instability thresholds are evaluated for a total
depth of $6\;mm$ in both configurations. The critical Marangoni curves exhibit  
 maxima at $M1$ and  $M2$
 as discussed above. In the case
Fc-75/water, hydrothermal waves which propagate from cold to hot
in the region located between $M1$ and $M2$ are found, for the
other values of the fractional bottom depth stationary
longitudinal rolls are forecast. The system acetonitrile/n-hexane
presents the same three regions as exhibited by the pair $5\; cSt$
silicon oil/HT-70 with a cod-2  point created by two oscillatory
modes with different frequency and wavenumbers. The general
features discussed earlier in this section remain valid  for all
the cases which have been examined.

Tuning the relative depths of both layers, according to our theoretical study  it would be possible to
obtain in the same experimental
configuration with a fixed pair of liquids all the patterns observed in different
single layer systems: from longitudinal stationary rolls to
hydrothermal waves with directions of propagation ranging from  small angles
of propagation (as in one-liquid systems with high $Pr$) to high angles (as in one-liquid systems with small $Pr$).

\section{Conclusions}\label{sec.conclusions}

We have investigated the onset of convection in systems constituted by
horizontal liquid layers subjected to a lateral heating.
 The system
is bounded by two rigid and thermally conductive horizontal walls
and the interface is supposed to be nondeformable. The acting
forces are the interfacial tension and the gravity. They are
cooperative in the lower layer and competitive in the upper one.
The number of parameters is greatly enhanced with respect to the
case of one-layer problems. We have considered three
different combinations of liquids, namely bilayers of
perfluorinate HT-70 and  $5\;cSt$ silicon oil, perfluorinate Fc-75
and water, and acetonitrile and n-hexane; these fluids have been
considered because they have already been the subject of
experiments involving a vertical heating \cite{Jue}. It is our hope that the
results of the present paper will motivate researchers to repeat
these experiments with a lateral heating.

 As soon as a horizontal temperature gradient is applied, a global circulation takes place in each layer.
The temperature and velocity fields of the basic state have been derived and
discussed. For a given experimental setup and a fixed heating,
 the two control parameters  are the total depth and the depth ratio between the liquids. In a diagram
representing the depth ratio versus the total depth, four regions
for the temperature and velocity profiles have been distinguished.
The main information drawn from this diagram can be summarized as
follows. The interaction between interfacial tension and buoyancy
can give rise  to: 1)  a counter-rotating cell in each layer, 2)
two counter-rotating cells in the upper liquid and one cell in the
lower one, in both cases one observes a flow from hot to cold near
the interface ; 3)  to two counter-rotating cells in the lower
liquid and a cell in the upper one, 4)  a counter-rotating cell in
each liquid, but with the flow being dragged near the interface
 from cold to hot in the last two cases. We have also determined the different
 temperature profiles that result from the  interplay between thermal diffusivity and fluid motion.

 The  linearized evolution equations for the perturbations of the basic state have been
established and the ensuing eigenvalue problem has been solved.
Bidimensional and three dimensional perturbations have been respectively considered.
A physical interpretation of the different regions found has been proposed.
The interest of the present analysis with respect a one-layer system is that it
exhibits a great variety of behaviors depending on the nature of the liquids the total
depth and relative depth between the layers.
Among the most interesting results can be mentioned the occurrence of a cod-2 point
arising as a consequence of the interaction between two Hopf modes with different frequency
and wavenumber.
 Furthermore, three different
patterns are predicted: 1) {\em hydrothermal waves} propagating
from cold to hot side with a small angle, which is typical of lateral heating in one-liquid
systems with high $Pr$,  2) {\em hydrothermal waves}  propagating
from hot to cold, and 3)  longitudinal {\em stationary rolls}.

The variety of results confirms that the problem of lateral heating in two-layer systems is a promising
and interesting area of research, both from the theoretical and experimental points of view.
It would be  specially interesting to study
the system close to  $M2$ when we have interplay
between two oscillatory modes.

\paragraph*{Acknowledgments}
  The authors wish to thank Mireille Dondlinger,  Thomas Desaive (Liege) and Jean Bragard (Pamplona)
for useful discussions.
This work has been supported  by the European Union through contract ICOPAC
HPRN-CT-2000-00136,  by the DGICYT (Spanish Government) grant   BFM2002-01002 and
the  PIUNA (Universidad de Navarra). One of us (SM) also acknowledges the  financial support
from "Asociaci\'on de Amigos de la Universidad de Navarra".

\nopagebreak[3]

\appendix
\section*{Appendix}

Expressions of the velocity and temperature profiles of the basic state in dimensionless units:

\begin{eqnarray}
 u_{0}^{(1)}&=&\frac{1}{48\,\left( a + \mu  \right) \,\nu }  \left[{-12\,a\,Ma\,\nu\left( 1 + 4\,z + 3\,z^2 \right)}\right. \label{eq:u11}\\
   &&+\left. {  Ra\,\left\{ a^3\,\alpha \,\mu  - a\,\nu   +
       4\,\mu \,\left( a^3\,\alpha  + \nu  \right)\,z  +
       3\,\left( a^3\,\alpha \,\mu  + 3\,a\,\nu  + 4\,\mu \,\nu  \right)\,z^2 + 8\,\left( a + \mu  \right) \,\nu \,z^3 \right\} } \right] \nonumber \\
u_{0}^{(2)}&=&\frac{1}{48\,a\,    \left( a + \mu  \right) \,\nu } \left[{ -12\,Ma\,\nu \left( a^2 - 4\,a\,z +
3\,z^2 \right)
   }\right. \label{eq:u22}\\
&&+ \left. {Ra\,\left\{ a^4\,\alpha \,\mu   - a^2\,\nu  +
       4\,a\,\left( a^3\,\alpha  + \nu  \right)\,z  -
       3\,\left( 4\,a^3\,\alpha  + 3\,a^2\,\alpha \,\mu  + \nu  \right)\,z^2 + 8\,a\,\alpha \,\left( a + \mu  \right)\,z^3  \right\} } \right] \nonumber\\
\tau^{(1)}&=&\frac{1}{2880\,    \left( a + \lambda  \right) \,\left( a + \mu  \right) \,\nu \,\kappa }\left[ {}\right. \label{eq:t11}\\
 && { 60\,Ma\,\left\{
       \,a^2\,\nu \,\left( a\,\lambda  + \kappa  \right)+
a\,\lambda \,\nu \,( a^2 - \kappa)\,z  -
       6\,a\,\left( a + \lambda  \right) \,\nu \,\kappa\,z^2    \right.  }\nonumber \\
&& {\left.  -    8\,a\,\left( a + \lambda  \right) \,\nu \,\kappa\,z^3 -3\,a\,\left( a + \lambda  \right) \,\nu \,\kappa\,z^4   \right\}    } \nonumber \\
&&{ +   Ra\,\left\{   a\,\nu \,\left( 5\,a^2\,\lambda  + 9\,a\,\kappa  + 4\,\mu \,\kappa  \right)  -
          a^4\,\alpha \,\left( 4\,a^2\,\lambda  + 9\,a\,\lambda \,\mu  + 5\,\mu \,\kappa  \right)  \right.     }\nonumber \\
&&{     - \left( \nu \,( -5\,a^3 + 9\,a\,\kappa  + 4\,\mu \,\kappa )  +
          \alpha \,( 4\,a^6 + 9\,a^5\,\mu  - 5\,a^3\,\mu \,\kappa ) \right)}\lambda \,z \nonumber \\
&&{+30\,a\,\left( a + \lambda  \right) \,\left( a^2\,\alpha \,\mu  - \nu  \right) \,\kappa\,z^2
 +40\,\left( a + \lambda  \right) \,\mu \,\left( a^3\,\alpha  + \nu  \right) \,\kappa\,z^3 } \nonumber \\
&&{ +      15\,\left( a + \lambda  \right) \,
        \left( a^3\,\alpha \,\mu  + 3\,a\,\nu  + 4\,\mu \,\nu  \right) \,\kappa\,z^4 +
24\,\left( a + \lambda  \right) \,\left( a + \mu  \right) \,\nu \,\kappa\,z^5 }\frac{}{}\left\}{} \right] \nonumber \\
\tau^{(2)}&=&\frac{-1}{2880\,a\,\left( a + \lambda  \right) \,\left( a + \mu  \right) \,\nu \,\kappa } \left[ {}\right. \label{eq:t22}\\
   && {\left.60 \,Ma\left\{  -\,a^3\,\nu \,\left( a\,\lambda  + \kappa  \right)+
          a^2\,\nu \,\left( -a^2 + \kappa  \right)\,z + 6\,a^2\,\left( a + \lambda  \right) \,\nu\,z^2  -
         8\,a\,\left( a + \lambda  \right) \,\nu\,z^3
          \right. \right. } \nonumber \\
&&{\left. + 3\,\left( a + \lambda  \right) \,\nu \,z^4        \right\}  +
      Ra\,\left\{ a^2\,\left( -\left( \nu \,\left( 5\,a^2\,\lambda  + 9\,a\,\kappa  + 4\,\mu \,\kappa  \right)
               \right)  + a^3\,\alpha \,\left( 4\,a^2\,\lambda  + 9\,a\,\lambda \,\mu  +
                5\,\mu \,\kappa  \right)  \right)  \right.  } \nonumber \\
&&   + a\left( \nu \,\left( -5\,a^3 + 9\,a\,\kappa  + 4\,\mu \,\kappa  \right)  +
             \alpha \,\left( 4\,a^6 + 9\,a^5\,\mu  - 5\,a^3\,\mu \,\kappa  \right)  \right)\,z\ -
         30\,a^2\,\left( a + \lambda  \right) \,\left( a^2\,\alpha \,\mu  - \nu  \right) \,z^2\nonumber \\
&&{          -40\,a\,\left( a + \lambda  \right)\left( a^3\,\alpha  + \nu  \right)\,z^3  +
         15\,\left( a + \lambda  \right)
           \left( a^2\,\alpha \,\left( 4\,a + 3\,\mu  \right)  + \nu  \right)z^4   -24\,a\,\alpha \,\left( a + \lambda  \right)\left( a + \mu  \right)z^5   } \left. {\left. \left.
           \right.
       \right\}} \right] \nonumber
\end{eqnarray}

\noi Note that the lower fluid (superscript 1) is defined in the interval $z\in[-1,0]$ and the upper fluid
(superscript 2) in the interval $z\in[0,a]$, according to the definitions given in section
\ref{sec:problemdefinition}.

\newpage

\newpage

\newpage

\newpage

\afterpage{\clearpage}

\newpage
\clearpage

\newpage
\clearpage

\newpage
\clearpage

\newpage
\clearpage

\newpage
\clearpage

\newpage
\clearpage

\newpage
\clearpage

\begin{center}
\begin{table}[t]
\begin{tabular}{l|lllllllll} \hline \hline
        & \multicolumn{1}{c}{$\rho$ } &  \multicolumn{1}{c}{$\nu$} &\multicolumn{1}{c}{$\lambda$} &\multicolumn{1}{c}{ $c_p$ }&\multicolumn{1}{c}{ $\alpha$ }&\multicolumn{1}{c}{$\gamma$}& \multicolumn{1}{c}{$Pr$}&\multicolumn{1}{c}{$Cr$} \\
Liquid & $(kg/m^{3})$ & $(10^{-6}\,m^{2}/s)$ & $(J/m\,s\,K)$&  $(J/kg\,K)$ & $(10^{-3}\,K^{-1})$& (N/m\,K) &  &  \\ \hline \hline
Silicon oil $5\;cSt$             & 920   & 5     & 0.117 & 1590  & 1.05  &                & 62.512&                \\
HT-70                            & 1680  & 0.5   & 0.07  & 962   & 1.1   &                & 11.54 &                \\
Silicon oil $5\;cSt$ /HT-70      & 0.548 & 10    & 1.671 & 1.653 & 0.954 & -7.3\,$\cdot10^{-5}$  &       & 2.1\,$\cdot10^{-6}$  \\ \hline
Water                            & 997   & 0.893 & 0.609 & 4180  & 0.257 &                & 6.111 &                \\
Fc-75                            & 1760  & 0.945 & 0.063 & 1046  & 1.4   &                & 27.397&                \\
Water/Fc-75                      & 0.566 & 0.945 & 9.59  & 3.996 & 0.183 &-4.7\,$\cdot10^{-5}$ &       & 3.3\,$\cdot10^{-7}$  
 \\ \hline
n-Hexane                         & 655   & 0.458 & 0.12  & 2270  & 1.41  &                & 5.675 &                \\
Acetonitrile                     & 776   & 0.476 & 0.118 & 2230  & 1.41  &                & 4.381 &                \\
n-Hexane/Acetonitrile            & 0.844 & 0.962 & 1.017 & 1.018 & 1     &-1\,$\cdot10^{-4}$     &       & 7.8\,$\cdot10^{-7}$ 
 \\ \hline \hline
\end{tabular}
 \caption{
Parameter values of the three pairs of liquids studied. The parameter ratio
  in each different configuration is also given. The Crispation number $Cr$ is given  for a depth of
$h^{(1)}=3\,mm$.
\label{propiedades}}
\end{table}
\end{center}

\newpage
\clearpage

\begin{center}
\begin{table}
\begin{eqnarray*}
 \begin{tabular}{c|c} \hline \hline
  $\nu$ & ${\hat h}_1$\\ \hline \hline
  0.1   & 84\%\\
   1    & 81\%\\
  10    & 68\%\\
\end{tabular}  &
\begin{tabular}{c|c} \hline \hline
  $\kappa$ & ${\hat h}_1$\\ \hline \hline
  0.1   & 90\%\\
   1    & 81\%\\
  10    & 62\%\\
\end{tabular}
\end{eqnarray*}
\caption{
Position of $M2$ as a function of the viscosity ratio (left) and thermal viscosity ratio
(right) of the two liquids. The case $\nu=1$, $\kappa=1$ corresponds to an ideal configuration formed by two layers of
the same fluid ($5\;cSt$ silicon oil). 
\label{nu-kappa}}
\end{table}
\end{center}

\end{document}